\documentclass[aps,pra,amsfonts,twocolumn]{revtex4}
\usepackage{epsfig}
\usepackage{bbm} 
\usepackage{multirow} 
\usepackage{lscape}
\usepackage{rotating}
\usepackage{hyperref}
\newcommand{\eq}{\begin{eqnarray}} 
\newcommand{\en}{\end{eqnarray}}
\def\bra#1{\mathinner{\langle{#1}|}}
\def\ket#1{\mathinner{|{#1}\rangle}}
\newcommand{\braket}[2]{\langle #1|#2\rangle}

\usepackage{ amssymb }
\newcommand{\eqs}{\begin{subequations}}
\newcommand{\ens}{\end{subequations}}
\usepackage{color}

\usepackage{placeins}
\usepackage{hyperref}
\usepackage{amsmath}

\begin{document}

\title{Dynamical effects of exchange symmetry breaking in mixtures of interacting bosons} 
\author{Malte C. Tichy} 
\author{Jacob F. Sherson} 
\author{Klaus M\o{}lmer} 
\address{Lundbeck Foundation Theoretical Center for Quantum System Research, Department of Physics and Astronomy, University of Aarhus, DK--8000 Aarhus C, Denmark}

\date{\today}
\begin{abstract}
In a double-well potential, a Bose-Einstein condensate exhibits Josephson oscillations or self-trapping, depending on its initial preparation and on the ratio of inter-particle interaction to inter-well tunneling. Here, we elucidate the role of the exchange symmetry for the dynamics with a mixture of two distinguishable species with identical physical properties, i.e.~which are governed by an \emph{isospecific} interaction and external potential. 
In the mean-field limit, the spatial population imbalance of the mixture can be described by the dynamics of a single species in an effective potential with modified properties or, equivalently, with an effective total particle number. The oscillation behavior can be tuned   by populating the  second species while maintaining the spatial population imbalance and all other parameters constant. In the corresponding many-body approach, the single-species description approximates the full counting statistics well also outside the realm of spin-coherent states. The method is extended to general Bose-Hubbard systems and to their classical mean-field limits, which suggests an effective single-species description of multicomponent Bose gases with weakly an-isospecific interactions. 
\end{abstract}

\maketitle

\section{Introduction}
Multi-component Bose gases \cite{Ho1996} exhibit a panoply of different quantum phases \cite{Roscilde2007,Buonsante2008a,Buonsante2009a}, which reflect the hierarchy of inter- to intra-species interaction parameters that compete with a possibly species-dependent \cite{Soltan-Panahi2011} external potential. The \emph{an-isospecificity}, i.e.~the distinct physical properties of the different species, is the main cause for the great increase of complexity that mixtures feature with respect to single-species Bose gases.

A minimalist exemplary system that contains many of the building blocks for quantum many-body dynamics is the bosonic Josephson junction \cite{Gati2007,Lee2012}, i.e.~a Bose-Einstein condensate (BEC) in a double-well potential. Despite its simplicity, it features qualitatively different dynamical regimes -- self-trapping and Josephson oscillations -- already in the single-component case \cite{Jack1996,Milburn1997,Smerzi1997,Raghavan1999,Gati2007}, which has been realized experimentally with coupled spatial modes \cite{Cataliotti2001,Albiez2005} and hyperfine states \cite{Zibold2010}. When a bi-component BEC is loaded into the double-well \cite{Ashhab2002,Sun2009,Naddeo2010,Julia-Diaz2009,Mele-Messeguer2011,Lee2004}, new features emerge: For example, chaos \cite{Ashhab2002,Xu2008}, phase separation transitions \cite{Zin2011} as well as ferromagnetic behavior \cite{Pu2001} can arise, relying on the distinct inter- and intra-species interactions in the system. 

If, on the other hand, all intra- and inter-species interactions are equal and all external potentials are species-independent, adding a second species to an initially homo-specific  system (or, equivalently, populating a second internal state) does, at first sight, not appear to change the system dynamics: The new species does not feature any distinct property in comparison to the present one. However, the mere assignment of a particle \emph{label} -- although invisible for the Hamiltonian -- breaks the exchange symmetry of the quantum state of the system, with important consequences.

Photons, for instance, do not interact, but indistinguishable photons  can interfere collectively. This leads to dramatic differences between the counting statistics of photons that can be distinguished by their polarization or frequency (and thus do not interfere collectively) and identical photons  \cite{Hong1987,Laloe2011,Mayer2011,Tichy2012}. The latter are governed by many-particle interference, such that, e.g.~events with many particles in one mode are privileged \cite{Mlmer1997,Mlmer2002}. 

\begin{figure*}[t]
\center
\includegraphics[width=\linewidth,angle=0]{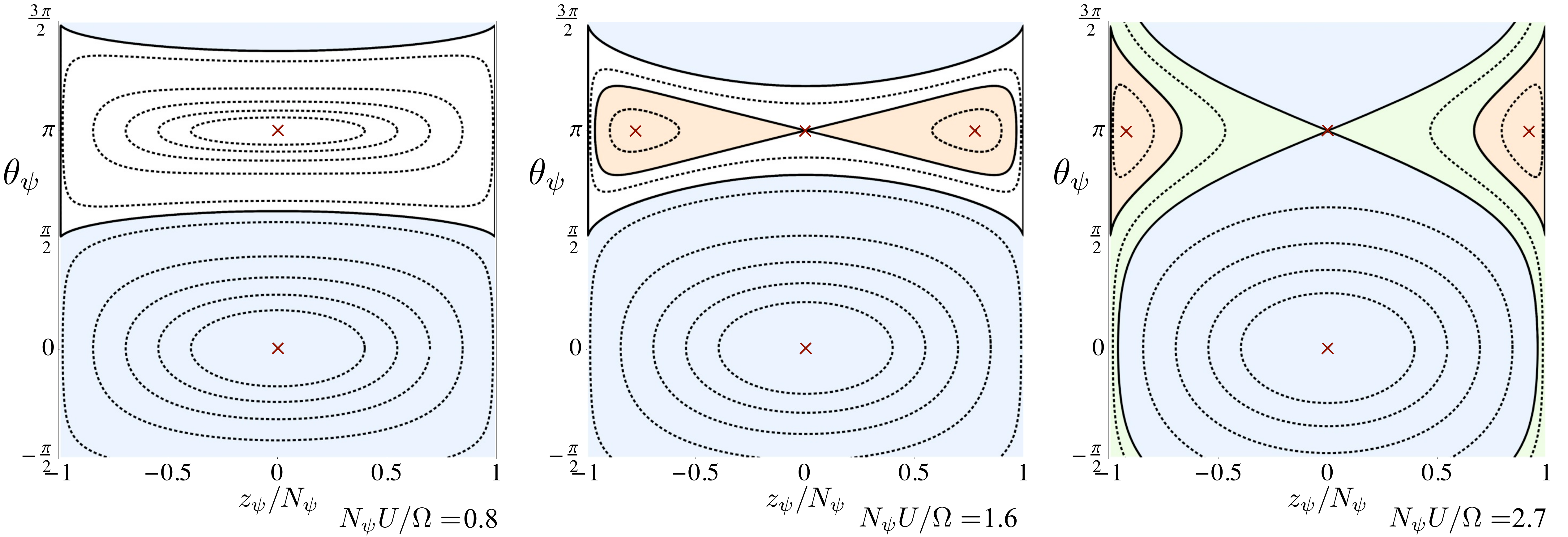} 
\caption{(color online) { Phase-space of the bosonic Josephson junction in the $z_\psi-\theta_\psi$-plane, for different values of $N_\psi U/\Omega$.} For convenience, we show the range $-\pi/2<\theta_\psi<3/2 \pi$. Thick solid lines denote the separatrices between the regions of qualitatively distinct behavior, dotted lines show exemplary trajectories, and fix points are denoted by red crosses. The fix point at $(z_\psi,\theta_\psi)=(0,\pi)$ ($N_\psi U /\Omega<1$) becomes unstable for $N_\psi  U/\Omega>1$. In this regime, two localized fixed points with $z_\psi\neq 0$ emerge. Solutions in the blue areas are characterized by $\langle z_\psi \rangle=\langle \theta_\psi \rangle =0$, solutions in the white areas lead to $\langle z_\psi \rangle=0, \langle \theta_\psi \rangle =\pi$. Self-trapped solutions are either $\pi$-modes (orange area, $\langle z_\psi \rangle \neq 0, \langle \theta_\psi \rangle =\pi$), or phase-running modes (green area, $\langle z_\psi \rangle \neq 0$, unbounded phase $\theta_\psi$). } 
 \label{phasespace}
\end{figure*}

For the bosonic Josephson junction, the question arises how the \emph{combination} of inter-particle interaction and a broken (or immaculate) exchange symmetry affects the dynamics. This investigation is the purpose of the present article: We compare a bi-species condensate with two species of identical properties, for which we provide an experimental protocol, to a single-species condensate. This reveals the consequences of broken exchange symmetry in a paradigmatic interacting many-particle system. We show that the population of a second species  can significantly affect the system dynamics: The tunneling behavior (self-trapping or oscillatory) can be switched by the manipulation of the species populations. This paves the road to the control of many-particle tunneling dynamics by merely populating different internal states. In the mean-field limit, the dynamics can be understood from the broken phase-coherence between the wells, and an exact single-species description is formulated. In the fully second-quantized formulation, the bi-species Fock-state counting statistics is efficiently approximated by the single-species model. We also investigate the relaxation of isospecificity, i.e.~species with slightly different parameters, and show that the single-species description still captures the essential dynamics within a wide range of parameters.  The generalization to larger systems therefore suggests an efficient description of weakly an-isospecific multicomponent Bose gases by a single species. 

We first review the main properties of the single-species bosonic Josephson junction -- in the discrete two-mode approach --  in Section \ref{GPET}. The physical consequences of the addition of a second species can be absorbed by an effective particle number or by an effective tunneling coupling, as shown in Section \ref{bispiciesdimer}. A complementary quantum calculation in the two-mode Bose-Hubbard model is performed for spin-coherent and  for Fock-states in Section \ref{BHaappp}, which confirms the validity of the approach beyond the mean-field limit. The generalization of the results to a general Bose-Hubbard Hamiltonian and to its mean field limit is presented in Section \ref{generalized}, before we conclude in Section \ref{Conclusions}.

\section{Single-species Gross-Pitaevskii equation and non-linear scaling}

\label{GPET}
In order to obtain analytic insight into the problem, we follow a two-mode treatment of the double-well potential, which is well-justified for very low temperatures and small condensate depletion \cite{Dalton2011}, and which was quantitatively verified experimentally in Ref.~\cite{Albiez2005}.  The dynamics of the macroscopic wavefunction that describes a (single-species) condensate in the symmetric double-well potential is then governed by the discrete Gross-Pitaevskii equation \cite{Smerzi1997,Milburn1997} (we set $\hbar=1$), 
\eq
i \frac{\partial }{\partial t} \psi_j &=& U  | \psi_j|^2 \psi_j - \frac \Omega 2 \psi_{3-j} , \label{GPE1} 
\en
where $j=1,2$; $\psi_1$ ($\psi_2$) is the amplitude of the wave-function in the left (right) well, such that $|\psi_j|^2$ is the expectation value of the number of particles in the $j$th well, and $N_\psi= |\psi_1|^2+|\psi_2|^2$ is the conserved total number of particles. The tunneling coupling $\Omega$ and the interaction  $U$ can be inferred from the overlap integrals of the exact localized wavefunctions of the double-well potential \cite{Smerzi1997}. The relevant parameter for the dynamics is the dimensionless quotient $N_\psi U /\Omega$, and we will measure all frequencies in units of $\Omega$. Since the scaling of $\Omega$ will also be discussed, we keep $\Omega$ as an explicit parameter. Eq.~(\ref{GPE1}) represents a self-trapping equation \cite{EILBECK1985,Scott1990}, which has been studied extensively \cite{Giovanazzi2000,Marino1999,Raghavan1999,Smerzi1997,Salgueiro2007,Ashhab2002,Dalton2011,Gati2007}. We review here the essential properties that will be useful for our analysis of the two-component case; in particular, we discuss a non-linear scaling property for the population imbalance between the wells. 

\subsection{Bloch-sphere description}
The dynamics of the system can be re-formulated as the motion of a vector $\vec v_\psi$ on a Bloch-sphere of radius $N_\psi$ \cite{Ashhab2002}, where 
\eq
\vec v_\psi = \left( \begin{array}{c} x_\psi \\ y_\psi \\ z_\psi \end{array} \right)
  =   \left( \begin{array}{c} 
2 ~\Re (\psi_1^\star \psi_2  )  \\
 2 ~\Im (\psi_1^\star \psi_2 )  \\
   |\psi_1|^2-|\psi_2|^2  
  \end{array} \right)  \label{zdef}  ,
\en
i.e.~the $z_\psi$-component of $\vec v_\psi$ is related to the population imbalance between the wells, while the relative phase between the left and right component of the condensate, 
\eq \theta_\psi = \arg(\psi_1^\star \psi_2), \en is encoded in

 \eq 
 x_\psi =\sqrt{N_\psi^2-z_\psi^2} \cos \theta_\psi, \ \ \ \   y_\psi = \sqrt{N_\psi^2-z_\psi^2}\sin \theta_\psi . \label{thetadif}
   \en  The equation of motion Eq.~(\ref{GPE1}) assumes the form of an optical Bloch equation with non-linear terms \cite{Lee2004},  
\eq 
\frac{\text{d}}{\text{d}t} \vec  v =\left( \begin{array}{ccc} 
- U ~y_\psi ~ z_\psi \\
\Omega~ z_\psi + U ~z_\psi ~x_\psi \\
-\Omega~ y_\psi  
 \end{array} \right)  , \label{blocheom}
\en
and the conserved energy reads
\eq H = \frac{U}{4} z_\psi^2  - \frac  \Omega 2 x_\psi \label{Hamilfull} . \en

Depending on the ratio $N_\psi U/\Omega$, the phase space of the system assumes a distinct structure \cite{Smerzi1997}. The three topologically distinct cases are shown in Fig.~\ref{phasespace}. For $N_\psi U/\Omega<1$ (left panel), in the Rabi regime, all solutions are oscillatory with $\langle z_\psi \rangle_t=0$. There are two elliptic fix points with $z_\psi=0$ and $x_\psi=\pm N_\psi$, i.e.~the average value of the phase $\theta_\psi$ is either $0$ or $\pi$. In the parameter range  $1<N_\psi U/\Omega<2$ (middle panel), the fix point $(x_\psi,y_\psi,z_\psi)=(N_\psi,0,0)$ remains stable, while $(x_\psi,y_\psi,z_\psi)=(-N_\psi,0,0) $ becomes unstable. Two new fix points  emerge, which are related to trapped $\pi$-mode oscillations with $\langle z_\psi \rangle_t \neq 0$ and $\langle \theta_\psi \rangle_t = \pi$. For $N_\psi U/\Omega>2$ (right panel), trapped phase-running modes emerge, for which the phase is unbounded. When the interaction is increased further, these phase-running modes dominate the picture more and more, whereas the oscillation region shrinks -- the Fock regime is attained.

\subsection{Second-order differential equation} \label{secondorderformulation}
The overall behavior of the system is governed by the phase-space structures shown in Fig.~\ref{phasespace}. In general, the self-trapping or oscillatory behavior of a solution depends on the initial preparation of the system and on the ratio $N_\psi U/\Omega$. If one focuses on the population imbalance $z_\psi$ only, a scaling symmetry relates solutions with different initial conditions and different values of $U$ and $\Omega$ to each other. For this purpose, we re-formulate the dynamics of $z_\psi$ by taking the time-derivative of Eq.~(\ref{blocheom}) \cite{Ashhab2002}, 
\eq 
\ddot z_\psi =- z_\psi( \Omega^2- 2 ~U H) - U^2 \frac{z_\psi^3}{2} , \label{closed}
\en
where $H$ is the value of the conserved energy, Eq.~(\ref{Hamilfull}). In terms of the initial preparation of $z_\psi$ and $\dot z_\psi$, Eq.~(\ref{closed}) reads 
\eq 
\ddot z_\psi  = \hspace{6cm}  \label{nd2order} \\
 - z_\psi \left( \frac{ U^2}{2} (z_\psi^2 - z_{\psi}(0)^2) + \Omega \left(\Omega + U x_\psi(0) \right) \right) , \nonumber
 \en
 where $x_\psi(0)$ can be expressed through $z_\psi(0)$ and $\dot z_\psi(0)$, except for its sign: 
 \eq 
x_\psi(0) =\pm \sqrt{N_\psi^2- z_{\psi}(0)^2 - \frac{\dot z_{\psi}(0)^2}{\Omega^2}} . \label{xpsizpsirelation}
 \en
Since the transformation $U \rightarrow -U$ is equivalent to $x_\psi \rightarrow -x_\psi$ \cite{Smerzi1997}, we choose here the convention $x_{\psi}(0)>0$ (positive square root) without loss of generality. This implies that the initial phase $\theta_\psi$ lies between $-\pi/2$ and $\pi/2$, see Eq.~(\ref{thetadif}). Any self-trapping then comes along with a phase-running mode.

\subsection{Non-linear scaling of population imbalance}
From the second-order differential equation of motion Eq.~(\ref{nd2order}) a scaling symmetry for the solutions can be derived by merely redefining the unit of time,
\eq 
z_\psi(N_\psi, z_{\psi}(0), \dot z_{\psi}(0), U, \Omega , t) \hspace{0.75cm} \nonumber \\
 =  z_\psi(N_\psi, z_{\psi}(0), \frac{\dot z_{\psi}(0)}{\alpha}, \frac U \alpha,  \frac \Omega \alpha , \alpha t) ,
 \en 
for any  positive scaling parameter $\alpha$. Also the  particle number $N_\psi$ can be scaled, 
\eq 
z_\psi(N_\psi, z_{\psi}(0), \dot z_{\psi}(0), U, \Omega , t) \nonumber \hspace{1.4cm}  \\  = \frac 1 \gamma z_\psi(\gamma  N_\psi, \gamma  z_{\psi}(0), \gamma  \dot z_{\psi}(0), \frac U \gamma , \Omega , t) , \label{scalepoptot}
\en
when the interaction strength $U$ is scaled reciprocally with $N_\psi$. Solutions on Bloch-spheres of different radii are thus related to each other.

For vanishing interaction, $U=0$, Eq.~(\ref{nd2order}) becomes a \emph{linear} differential equation, and 
\eq 
z_\psi(N_\psi, z_{\psi}(0), 0, U=0, \Omega , t) \nonumber  \hspace{0.85cm} \\  = \frac 1 \beta z_\psi(N_\psi, \beta  z_{\psi}(0), 0, U=0, \Omega , t) , \label{linearamplit}
\en
where $0<\beta<N_\psi/z_{\psi}(0) $ is a real parameter that scales the initial population imbalance,  keeping the Bloch-sphere radius constant.  Rabi oscillations of different amplitude are related to each other, and the scaling Eq.~(\ref{linearamplit}) reflects the trivial topology of the phase space for  $U=0$, where all solutions $z_\psi(t)$ are sinusoidal.

In general, a finite non-linearity $U\neq 0$ breaks the linear scaling property Eq.~(\ref{linearamplit}): Different initial conditions for the population imbalance $z_{\psi}(0)$ lead to qualitatively different solutions, as already apparent from the phase-space structure for $N_\psi U/\Omega>1$ in Fig.~\ref{phasespace}. Eq.~(\ref{nd2order}), however, allows to relate solutions with different initial population imbalance to each other. More precisely, for vanishing initial derivative, $\dot z(0)=0$, a solution with $z_\psi(0)=z_{0}$ can be related to a scaled solution with $z_\psi(0)=\zeta_0$, such that 
\eq 
z_\psi(N_\psi, z_{0}, 0, U, \Omega , t) = 
\frac {z_0}{\zeta_{0}} z_\psi(N_\psi, \zeta_{0}, 0, \widetilde U,\widetilde  \Omega , t) ,
\label{scalingprop}
\en
when the interaction $U$ and the tunneling coupling $\Omega$ are scaled according to
\eqs \label{scalingpfor}
\eq 
\widetilde U& =&  U \left| \frac{z_0}{\zeta_0} \right|  \label{forU}  , \\
\widetilde \Omega &=&  \sqrt{\Omega \left( \Omega + U \sqrt{N_\psi^2-z_0^2} \right) + \frac{U^2 ~z_0^2}{4~\zeta_0^2} \left(N_\psi^2-\zeta_0^2 \right)} \nonumber \\  && - \frac {U}{2}  \left| \frac{z_0}{\zeta_0} \right|  \sqrt{N_\psi^2-\zeta_0^2}  \label{forom} ,
\en
\ens
remember that we assume $\theta_\psi(0)=0$, or, equivalently, $x_\psi(0)>0, y_\psi(0)=0$. A change of the initial population imbalance $z_\psi(0)$ from $z_0$ to $\zeta_0$ can thus be compensated by choosing a scaled interaction $\widetilde U$ and tunneling $\widetilde \Omega$. The relative phases of the two solutions, $\theta_\psi$, or, equivalently, the motion of $x_\psi$ and $y_\psi$ are, however, not linearly related to each other. Therefore, this scaling property is not apparent from the phase-space picture, since it relates solutions with possibly different topology to each other, and $ U/\Omega \neq \widetilde U / \widetilde \Omega$, in general. Note that $\tilde \Omega$ may become complex. 

The scaling of the interaction $U$, explicit in Eq.~(\ref{forU}), can be understood rather intuitively: When we relate a solution with, say, $z_\psi(0)=z_0=N_\psi$ (all particles in the left well) to a solution with $z_\psi(0)=\zeta_0<N_\psi$, the effect of the interaction $U$ will be weaker for the system with smaller initial imbalance $\zeta_0$. To retrieve the original behavior of the solution with $z(0)=N_\psi$, this effect has to be counter-balanced by the upscaling of $U$.

The scaling of the frequency $\Omega$ is more intricate. In the first place, $\widetilde \Omega \neq \Omega$ only when interaction is present.  Assuming $\zeta_0<z_0$, such smaller initial population imbalance leads to the same effect as a larger (smaller) tunneling  $\widetilde \Omega> \Omega$ ($\widetilde \Omega < \Omega$), for repulsive (attractive) interaction $U>0$ ($U<0$). 

\begin{figure}
\center
\includegraphics[width=.8\linewidth,angle=0]{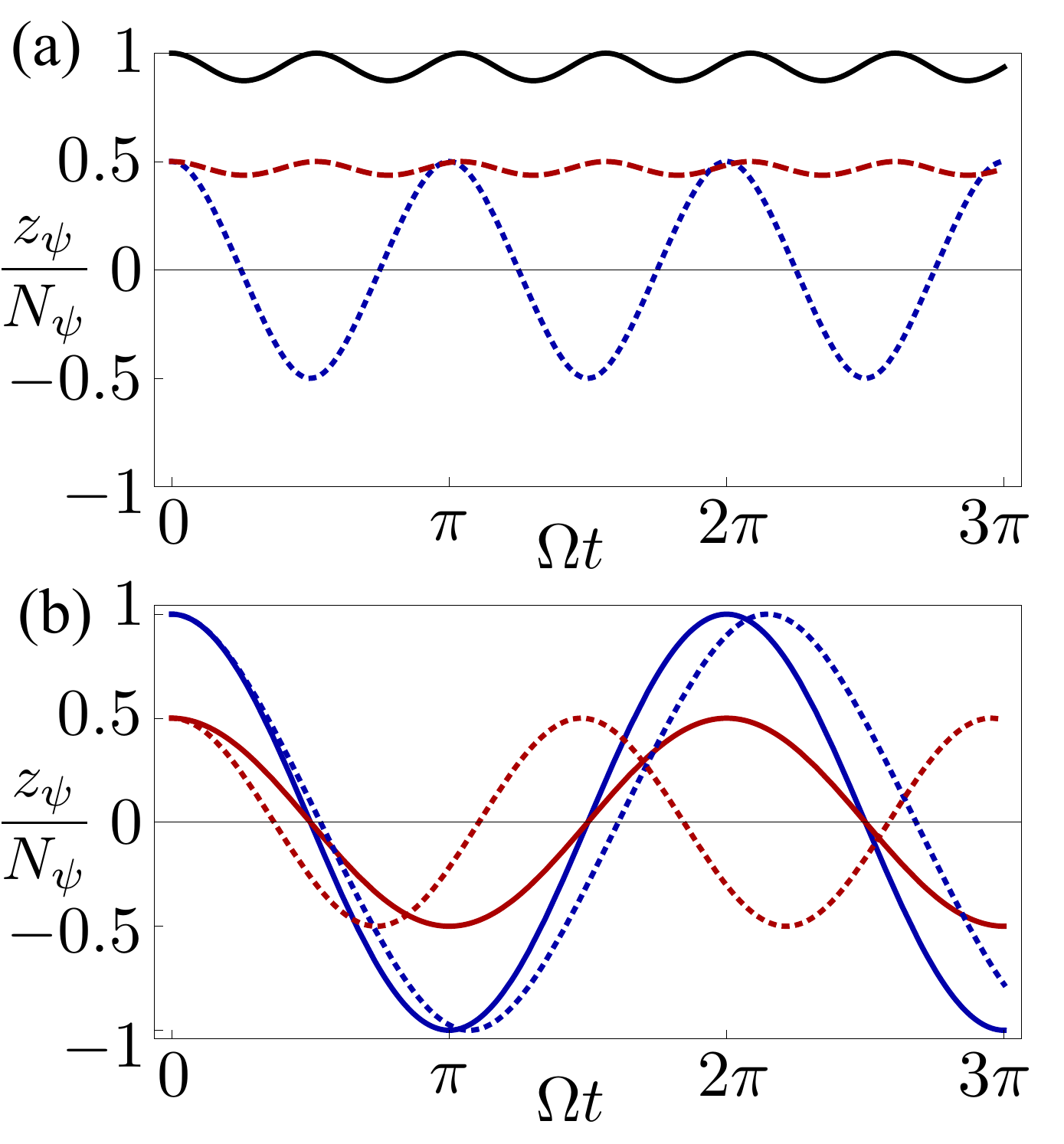} 
\caption{(color online) { Population imbalance scaling. }  (a) For $N U/\Omega=4.1$, the initial preparation $z_\psi(0)/N_\psi=1$ leads to self-trapping (black solid line), while for $z_\psi(0)/N_\psi=0.5$, oscillations emerge (blue dotted line). According to the scaling property Eq.~(\ref{scalingpfor}), the parameters $N_\psi \widetilde U/\Omega={8.2}$, $\widetilde \Omega =0.13813$, $z_\psi(0)/N_\psi=0.5 $ recover the qualitative behavior of the solution $z_\psi(0)/N_\psi=1$ (red dashed line). (b) For $z_\psi(0)/N_\psi=1$, increasing the interaction from $N_\psi U/\Omega=0$ (solid blue) to $N_\psi U/\Omega=1$ (dotted blue) leads to a decrease of the observed frequency, whereas for $z_\psi(0)/N_\psi=0.5$, the opposite behavior is observed, which can be understood via the scaling property.} 
 \label{scalingprops}
\end{figure}

The scaling property is illustrated in Fig.~\ref{scalingprops}: In (a), for a constant interaction $N_\psi U/\Omega =4.1$, the solution for $z_\psi(0)/N_\psi=1$ is self-trapped, whereas for $z_\psi(0)/N_\psi=1/2$, it oscillates. The scaled parameters $\widetilde U$, $\widetilde \Omega$ allow us to recover the self-trapped behavior at a scaled population imbalance. In particular, the scaling properties allow us to understand why a small repulsive interaction leads to slower oscillations for $z_\psi(0)=1$, while it leads to faster oscillations when $z(0)<1$, as shown in Fig.~\ref{scalingprops}(b).

\section{Bi-species dimer}  \label{bispiciesdimer}
Having established the behavior of a single-component BEC in the double-well, we now add a second component with very similar physical properties. The dynamics of the second species, which we will refer to as $\phi$, is governed by the same physical parameters as the first species $\psi$;  the tunneling rates are equal, $\Omega:= \Omega_\psi=\Omega_\phi$, and the intra- and inter-species interaction strengths fulfill 
\eq 
U_{\psi,\phi}=r U_{\psi,\psi} = r U_{\phi,\phi}=r U.  \label{isospecificityU}
\en
We will assume $r=1$, i.e.~perfect isospecificity, unless indicated explicitly.  A realistic physical implementation of the model consists in the second species being realized by a different hyperfine-state, i.e.~one can convert particles between the species at will, by applying suitable Rabi pulses on one or both wells. We are interested in the dynamics of the spatial population imbalance when the second species $\phi$ is populated at the expense of $\psi$, i.e.~for a constant total particle number $N=N_\psi+N_\phi$ and constant initial spatial population imbalance $z_{\text{tot}}=z_\psi+z_\phi$. 

\subsection{Equations of motion and conserved quantities}
In analogy to Eq.~(\ref{GPE1}), the coupled discrete Gross-Pitaevskii-equations for the bi-species system read 
\eqs \label{eom2}
\eq 
i \frac{\partial }{\partial t} \psi_j &=&  U \left( | \psi_j|^2 + r |\phi_j|^2 \right) \psi_j - \frac {\Omega}2 \psi_{3-j} ,  \label{eom21} \\
i \frac{\partial }{\partial t} \phi_k &=&  U \left(r | \psi_k|^2 + |\phi_k|^2 \right) \phi_k - \frac \Omega 2 \phi_{3-k} , \label{eom22} 
\en
\ens
where $k,j=1,2$ and we assume $r=1$. The coupling between the two species occurs via the non-linear interaction term containing $U$. In the following, the total particle number of each species is conserved, 
\eq 
|\psi_1(t)|^2+|\psi_2(t)|^2 = N_\psi,  \label{alphaconser} \    |\phi_1(t)|^2+|\phi_2(t)|^2 = N_\phi ,
\en
for all times. The two Bloch-vectors $\vec v_\psi$ and $\vec v_\phi$ that describe the two species, defined analogously to Eq.~(\ref{zdef}), evolve on spheres of radius $N_\psi$ and $N_\phi$, respectively. The coupled equations of motion Eq.~(\ref{eom2}) can be re-formulated for the Bloch-vectors $\vec v_\psi$ and $\vec v_\phi$ to give
\eq
\frac{\text{d}}{\text{d}t} \vec  v_\beta & = &\left( \begin{array}{ccc} 
- U ~y_\beta ~ (z_\psi+z_\phi) \\
\Omega~ z_\beta + U ~x_\beta ~ (z_\psi+z_\phi) \\
- \Omega~ y_\beta  
 \end{array} \right)  ,     \label{blocheom2a} 
\en
where $\beta=\psi, \phi$, while the total energy
\eq 
H_{\text{tot}} &=&  \frac U 4 (z_\psi + z_\phi)^2 - \frac \Omega 2 \left( x_\psi + x_\phi \right) \nonumber \\ &=& 
\frac U 4 z_{\text{tot}} ^2 - \frac \Omega 2  x_{\text{tot}}    . \label{twocomphamil}
\en
is conserved. Since the total Hamiltonian of the system depends only on the spatial population imbalance, $z_{\text{tot}}$, and on the sum of the $x$-components of the Bloch-vectors, $x_{\text{tot}}$, it assumes the same form as for a single species, see Eq.~(\ref{Hamilfull}). Consequently, the equations of motion for the \emph{total} Bloch-vector $\vec v_{\text{tot}}$,
\eq 
\vec v_{\text{tot}} = \vec v_\psi+\vec v_\phi  \label{totalblochv}, 
\en
 possess the same form as for the Bloch-vector of a single species, Eq.~(\ref{blocheom}) \cite{Ashhab2002}. The length of $\vec v_{\text{tot}}$ 
 is a constant of motion, which is  a consequence of the isospecificity of the dynamics, i.e.~of $U_{\psi,\psi}=U_{\phi,\psi}=U_{\phi,\phi}$ and $\Omega_\phi=\Omega_\psi$. Note that while the dynamics of the total population imbalance is integrable, this is not necessarily the case for the dynamics of the individual species, i.e.~the difference between the imbalances of the two species, $\vec v_{\text{rel}} =\vec v_\psi - \vec v_\phi$, eludes a closed solution  \cite{Xu2008}. 
  When the interactions are not isospecific ($r\neq 1$, $U_{\psi,\phi}\neq  U_{\phi,\phi}$), $|\vec v_\text{tot}|$ is not conserved anymore. Similarly to the triple-well system \cite{Franzosi1,Franzosi2,Liu2007}, chaos can then emerge in the two-species double-well-system \cite{Ashhab2002,Xu2008}.  

The dynamics of the mixture can be computed in two different ways. We can either integrate the equations of motion of the individual species, Eq.~(\ref{blocheom2a}), or we can express the dynamics in terms of a single effective species, as suggested by the total Hamiltonian Eq.~(\ref{twocomphamil}). In order to illustrate the underlying physical mechanism, we will start with the former and return to the latter in the following sections. 

\subsection{Dynamical effects of a second species} \label{effectsofasecond}
\begin{figure*}
\center
\includegraphics[width=\linewidth,angle=0]{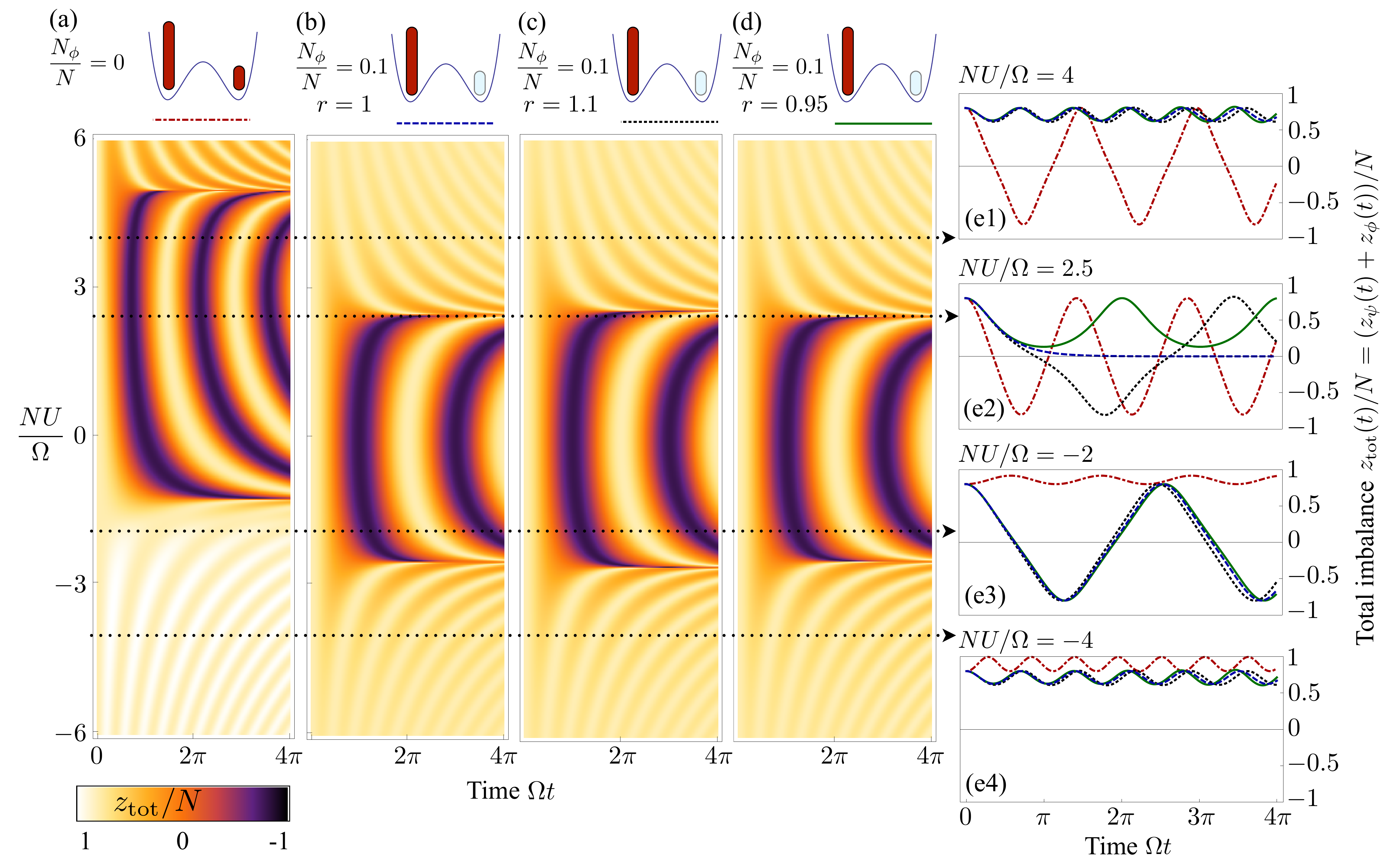} 
\caption{(color online)  Total population imbalance dynamics for (a) one species, (b) two perfectly isospecific species, (c,d) two slightly an-isospecific species with a ratio of the inter-to-intra-species interaction strength $r=1.1$ (c), and $r=0.95$ (d), i.e.~$U_{\psi,\phi}= r U$, see Eq.~(\ref{isospecificityU}).  The initial imbalance is $z_{\text{tot}}(0)/N=0.8$ always, and only the population of the right well is modified. In (a), the second species is absent, in (b,c,d),  the population in the right well fully appertains to the second species (see sketches above (a,b,c,d)). The panels (e1)-(e4) on the right show examples for the four situations and selected values of $N U/\Omega $  (dot-dashed red: (a), dashed blue: (b), dotted black: (c), solid green: (d)).}  
 \label{MainOverview}
\end{figure*}

The integration of the equations of motion, Eq.~(\ref{blocheom2a}), allows us to directly compare situations with one and two species.  The dynamics of the spatial population imbalance $z_{\text{tot}}(t)$ of a single-species and a fully isospecific bi-species BEC is shown in Fig.~\ref{MainOverview} (a,b), the  behavior of weakly an-isospecific mixtures can be observed in (c,d), we will discuss the relaxation from isospecificity in Section \ref{robustness1} below. The initial spatial population imbalance is set to $z_{\text{tot}}(0)/N=0.8$, while we vary the population of the second species and the interaction $U$  (see also the sketches above the graphs). We keep the total population $N$ constant and an initially vanishing phase between the wells, for both species, see also Fig.~\ref{blochvectorcouplingfig}. We assume that the two species correspond to different ground states of the same atoms and that a Rabi-pulse is applied on the right well, which transfers population from the first to the second species. In Fig.~\ref{MainOverview}(a), no Rabi pulse is applied and only one species is present in the system, $N_\phi/N=0$. In (b), a $\pi$ pulse transfers all the right-well population to the second species, $N_\phi/N=0.1$. For intermediate values of $N U/\Omega$ ($1\lesssim |N U /\Omega| \lesssim 6 $), populating the second species changes the dynamics of the spatial population imbalance $z_{\text{tot}}(t)$ strongly, the location of the oscillation regime on the $N U/\Omega$-axis is shifted to smaller values for increasing population of the second species. Given a fixed interaction parameter $N U/\Omega$, the behavior can therefore qualitatively differ for the different populations: In panel (e1), for $N U/\Omega=4$, the single-species BEC is clearly in the oscillation regime, while for $N_\phi/N=0.1$, no Josephson-like oscillations are observed anymore, the system is self-trapped. For $N U/\Omega=-2$, the opposite behavior is observed: While for $N_\phi/N=0.1$, the system oscillates, it is self-trapped for the other scenario.

We can understand the impact of the second species from the structure of the Hamiltonian Eq.~(\ref{twocomphamil}): Let us first consider repulsive interactions, i.e.~$U>0$. Since we assumed $x_{\text{tot}}(0)>0$,  the phase-coherence part of the Hamiltonian, $-\Omega x_{\text{tot}}(0)/2$, is initially  negative, which favors oscillations. The total energy, $H_{\text{tot}}=U z_{\text{tot}}(0)^2/4-\Omega x_{\text{tot}}(0)/2 $, is reduced -- for $|H_{\text{tot}}|<\Omega/2$, the system is in the oscillation regime, since $z_{\text{tot}}=0$ can then be attained energetically. In other words, initial coherence between the two wells with $x_{\text{tot}}(0)>0$ favors oscillations. Populating the second species in one well partially destroys the coherence, which increases the total energy and pushes the system towards self-trapping. 

Attractive interaction ($U<0$) leads to the opposite effect: The interaction term in the Hamiltonian is negative, and the phase-coherence part $-\Omega x_{\text{tot}}(0)/2$ is beneficial for self-trapping. Destroying the phase coherence by populating a second species is beneficial for Josephson oscillations. In general, the competition of interaction and phase-coherence in the Hamiltonian for the initial states in (a) breaks the symmetry $U \rightarrow -U$ present in panel (b). 

Summarizing, the behavior of the system can be changed from self-trapping to Josephson oscillations by the population of a second species, without changing $z_{\text{tot}}(0)/N$, $U$ or $\Omega$. As can be seen, the behavior of the two-species mixture and of the single-species condensate are similar -- although subject to a shift on the $N U/\Omega$-axis.

\subsection{Effective tunneling coupling}
The effect of a second species on the mean field dynamics can be modeled by a single species system with an effective tunneling coupling $\bar \Omega$. To derive this in analogy to Section \ref{secondorderformulation}, we consider the dynamics of the spatial population imbalance $z_{\text{tot}}$. Similarly to Eq.~(\ref{nd2order}), we find a second-order differential equation with parameters that depend on the initial conditions: 
\eq 
\ddot z_{\text{tot}} =  \label{ztotdiffequ} \hspace{6.5cm} \\ 
- z_{\text{tot}}\left(\frac{ U^2 }{2}(z_{\text{tot}}^2 - z_{\text{tot}}(0)^2  ) +   \Omega \left(\Omega + U x_{\text{tot}}(0)  \right) \right),  \nonumber 
\en
where $x_{\text{tot}}(0)$ and $z_{\text{tot}}(0)$ are defined through Eq.~(\ref{totalblochv}).  $x_\psi(0)$ can be related to $z_\psi(0)$ and $\dot z_\psi(0)$ through Eq.~(\ref{xpsizpsirelation}), which is valid analogously for $x_\phi(0)$, $z_\phi(0)$ and $\dot z_\phi(0)$.  We assume in the following $\dot z_{\psi}(0)=\dot z_{\phi}(0)=0$ and $x_{\psi}(0),x_{\phi}(0) \ge 0$. By comparison to Eq.~(\ref{nd2order}), we find that the spatial population imbalance for the two-species condensate, $z_{\text{tot}}(t)$, exhibits the same time-dependence as the imbalance of a single-species BEC that is prepared with the same initial imbalance, $z_\psi(0)=z_{\text{tot}}(0)$, but which experiences an effective tunneling coupling $\bar \Omega$, where 
\begin{widetext}
\eq 
\bar \Omega= \left[ \Omega\left( \Omega + U \left(\sqrt{N_\psi^2-z_{\psi}(0)^2}+\sqrt{N_\phi^2-z_\phi(0)^2}
\right) \right) +\frac{U^2 \left(N^2-{z_{\text{tot}}(0)}^2\right)}{4}  \right]^{1/2}  -\frac U 2 \sqrt{N^2-{z_{\text{tot}}(0)}^2}  \label{rescaledOme} ,
\en
\end{widetext}
is found by solving the resulting quadratic equation. Formally, $z_{\text{tot}}(t)$  satisfies
\eq
z_{\text{tot}}(N_\psi, N_\phi, z_{\psi}(0), z_{\phi}(0), U , \Omega,t ) \nonumber \\ 
=  z_\psi(N_\psi+N_\phi,  z_{\psi}(0)+z_{\phi}(0) , U ,\bar \Omega, t) , \label{effectivecoupling}
 \en
 where $z_\psi(t)$ is the solution to Eq.~(\ref{closed}).  In other words, the effect of breaking the phase coherence due to the addition of a second species is equivalent to a change of the tunneling coupling to $\bar \Omega$. Consistently with the findings discussed by means of Fig.~\ref{MainOverview}, Eq.~(\ref{rescaledOme}) shows that $\bar \Omega \le \Omega$ for repulsive interactions $U>0$, while $\bar \Omega \ge \Omega$ for attractive $U<0$.

\subsection{Effective particle number} \label{amplitudescaling}
The incorporation of the effect of the second species in the modified tunneling coupling parameter $\bar \Omega$ in Eq.~(\ref{rescaledOme}) relies on the exact form of the two-mode differential equations Eqs.~(\ref{nd2order}) and (\ref{ztotdiffequ}). A more general way to understand the dynamics of $z_{\text{tot}}$, which will also be applicable in larger systems, can be found by considering the emerging total Bloch-vector. In the bi-species case, the length of the Bloch-vector of either species reflects the respective population,  $ 
|\vec v_{\psi}|=N_\psi,  \  |\vec v_{\phi}|=N_\phi . $ 
The total number of particles in the bi-species case is the sum of the lengths of the two Bloch-vectors, $  |\vec v_{\psi}|+|\vec v_{\phi}|=N_\psi+N_\phi=N$, while the length of the total Bloch-vector $\vec v_{\text{tot}}$, defined in Eq.~(\ref{totalblochv}), is, in general, smaller than $N$.

The dynamics of  a bi-species condensate of a total particle number $N$ is thus described by the Bloch-vector $\vec v_{\text{tot}}$ that evolves on a sphere of radius $N_{\text{eff}}$, 
\eq 
N_{\text{eff}} &=&|\vec v_{\text{tot}}|= \sqrt{ |\vec v_\psi + \vec v_\phi |^2 } \nonumber \\ & =&\sqrt{ N_\psi^2 + N_\phi^2+ 2 \vec v_\phi \vec v_\psi  } , \label{classicaleffpn}
\en
i.e.~the system behaves like a single-species system with a \emph{reduced} total population $N_{\text{eff}}=|\vec v_{\text{tot}}|$. 

Formally, the spatial population imbalance in the bi-species case fulfills, 
\eq
z_{\text{tot}}(N_\psi, N_\phi, z_{\psi}(0), z_{\phi}(0), \Omega, U, t) \nonumber  \\
= z_\psi( N_{\text{eff}},  z_{\psi}(0)+ z_{\phi}(0), \Omega, U, t) , \label{effpan}
\en
where $N_\text{eff}$ is given in Eq.~(\ref{classicaleffpn}), which should be compared with Eq.~(\ref{effectivecoupling}), where we found an effective \emph{tunneling coupling} $\bar \Omega$. Applying the scaling of the Bloch-sphere radius, Eq.~(\ref{scalepoptot}), and of the initial population imbalance, Eq.~(\ref{scalingprop}), we can reconcile the two perspectives (Eqs.~(\ref{effectivecoupling}) and (\ref{effpan})) with each other, 
\eq
z_\psi\left( N_{\text{eff}},  z_{\psi}(0)+ z_{\phi}(0), \Omega, U, t \right) \hspace{2.3cm}   \nonumber \\
\overset{(\ref{scalepoptot})}{=} \frac{N_{\text{eff}}}{N} z_\psi \left( N,  \frac{N}{N_{\text{eff}}} \left( z_{\psi}(0)+ z_{\phi}(0) \right) , \Omega, \frac{N_{\text{eff}}}{N} U, t\right) \nonumber \\ 
\overset{(\ref{scalingprop})}{=} z_\psi\left( N, \left( z_{\psi}(0)+ z_{\phi}(0) \right)  ,\widetilde \Omega,  U, t\right) , \hspace{2.1cm} 
\en
where $\widetilde \Omega$ is given by Eq.~(\ref{forom}). Inserting the parameters confirms that the effective particle number $N_{\text{eff}}$ indeed leads to an effective frequency $\widetilde \Omega$ with $\widetilde \Omega= \bar \Omega$,  consistent with Eq.~(\ref{rescaledOme}).

\begin{figure}
\center
\includegraphics[width=\linewidth,angle=0]{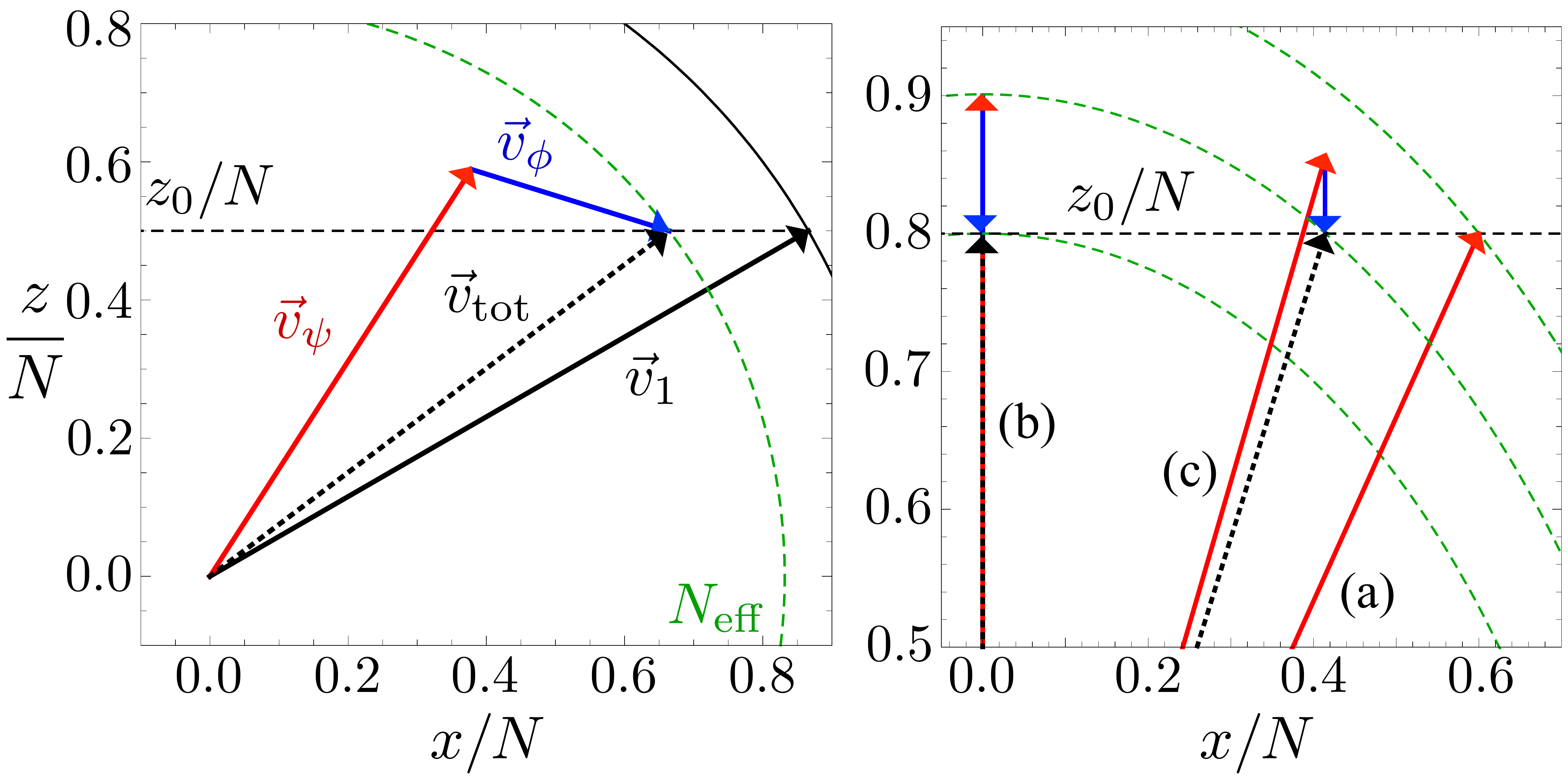} 
\caption{(color online) { Coupling of the Bloch-vectors that describe the two species. } Left panel: $\vec v_\phi$ (blue) and $\vec v_\psi$ (red) couple to a total Bloch-vector $\vec v_{\text{tot}}$ (black, dotted), as compared to a single-species Bloch-vector $\vec v_1$ (black, solid) of length $|v_\psi|+|v_\phi|=N_\psi+N_\phi=N$, with identical  $z$-component $z_0$. The length of $\vec v_{\text{tot}}$ (dashed green Bloch-sphere) is reduced in comparison to the single-species case (thin solid black Bloch-sphere). Right panel: Depending on the relative orientation of $\vec v_\psi$ and $\vec v_\phi$, the resulting effective vector length $|\vec v_{\text{tot}}|$ varies. The configurations (a,b) correspond to the respective physical situations shown in Fig.~\ref{MainOverview}: Only one species is present in (a); for (b) $|\vec v_\phi|/N=0.1$ and $\vec v_\psi$ and $\vec v_\phi$ are anti-parallel, while in the intermediate case (c), $|\vec v_\phi|/N=0.05$ (not shown in Fig.~\ref{MainOverview}). }  
 \label{blochvectorcouplingfig}
\end{figure}

The geometric coupling of the Bloch-vectors allows us to understand the effect of concerting or obstructing relative phases. When the relative phases of both species and the relative population imbalances are equal, $\theta_\phi(0)=\theta_\psi(0)$, $z_\psi(0)/N_\psi=z_\phi(0)/N_\phi$, the dynamics of the two species are concerted, and the collective motion corresponds to the one of a single species with the same phase difference. Geometrically speaking, the Bloch-vectors $\vec v_\psi$ and $\vec v_\phi$ are parallel, and the mixture behaves like a single species with $N=N_\psi+N_\phi$. Likewise, $z_\phi/N_\phi=z_\psi/N_\psi$ implies $\bar \Omega=\Omega$ in Eq.~(\ref{rescaledOme}).

When $\theta_\phi(0) \neq\theta_\psi(0)$ or $z_\psi(0)/N_\psi \neq z_\phi(0)/N_\phi$ the different initial phases or imbalances work against each other, and the motion of $\psi$ is then not concerted with the motion of $\phi$. The dynamics can then be described by a single species with smaller population.  Geometrically speaking, depending on their relative orientation, the length of the total Bloch-vector can be close to $N_\psi+N_\phi$, or be significantly smaller, as illustrated in Fig.~\ref{blochvectorcouplingfig}. As an example, we return to the configurations considered in Fig.~\ref{MainOverview}, for which the coupling of the vectors is depicted in the right panel of Fig.~\ref{blochvectorcouplingfig}. In both (b) and (c) the second species is initially localized in the right well and, consequently, $z_\phi=-N_\phi$, the Bloch-vector points down. As the population of the second species is increased, $\vec v_\psi$  is tilted towards the $z$-axis and the length of the total Bloch-vector decreases.

In general, the dynamics of the system is invariant under the simultaneous application of a Rabi pulse on both wells, which geometrically corresponds to a splitting of the constant total Bloch-vector $\vec v_{\psi}+\vec v_\phi=\vec v_\text{tot}$ into two new parts $\vec v^\prime_\psi$ and $\vec v^\prime_\phi$, with $\vec v^\prime_{\psi}+\vec v^\prime_\phi=\vec v_\text{tot}$. 

In particular, one does not alter the system behavior when applying a Rabi pulse if only one well is occupied initially such that $z_{\text{tot}}(0)=\pm N$. No phase coherence is broken by the Rabi pulse, since no phase relationship between condensate wavefunctions in the left and right well is present in the first place. 

\subsection{Robustness against deviations from isospecificity}  \label{robustness1}
 Since conventional optical traps operate with linearly polarised light that is far detuned from atomic resonance frequencies, atoms in different hyperfine states typically experience the very same optical potential, which motivates the species-independent tunneling rate $\Omega$. However, the species may be subject to slightly different inter- and intra-species scattering lengths \cite{Burke1997,Matthews1998,Mele-Messeguer2011}, i.e.~$r$ may differ from unity (see Eq.~(\ref{isospecificityU})). The question arises how deviations from the ideal parameters impact  on the system dynamics, since the treatment above  relies heavily on the conservation of $|\vec v_{\text{tot}}|$. Relaxing the assumption $r=1$, 
 the Hamiltonian becomes explicitly dependent on $\vec v_{\text{rel}}=\vec v_\psi - \vec v_\phi $, 
\eq 
H&=&\frac{U}{4}\left(\frac{(1+r)}{2} z_{\text{tot}}^2 + \frac{(1-r)}{2} z_{\text{rel}}^2 \right)- \frac \Omega 2 x_{\text{tot}} ,
\en
and the dynamics of the total Bloch vector also depends on the relative vector $\vec v_{\text{rel}}$,
\eq
\frac{\text{d} \vec v_{\text{tot}}}{\text{d}t}  = \left(\begin{tabular}{c}
$\frac U 2 \left(\left( {r-1} \right) y_{\text{rel}} z_{\text{rel}} -\left({r+1} \right)  y_{\text{tot}} z_{\text{tot}}\right) $ \\
$\frac U 2  \left(\left( {r+1} \right)  x_{\text{tot}} z_{\text{tot}} -\left( {r-1}{} \right) x_{\text{rel}} z_{\text{rel}}  \right) + \Omega z_{\text{tot}} $ \\
$-\Omega y_{\text{tot}} $
 \end{tabular} \right) . \label{totalvectordeo}
 \en
The length of the total Bloch-vector, $|\vec v_{\text{tot}}|$ is therefore not conserved anymore and  the time-derivative of its square reads
\eq 
\frac{\text{d}}{\text{d}t}  |\vec v_{\text{tot}}|^2 = 2(1-r) U (x_\psi y_\phi - x_\phi y_\psi ) z_{\text{rel}} .
\en
Thus, due to the different intra- and inter-species interaction strengths, the relative Bloch vector $\vec v_{\text{rel}}$ perturbs the motion of the total Bloch vector $\vec v_{\text{tot}}$, which  experiences an effective particle number that varies in time, or, equivalently, a varying interaction strength. 

The influence of the relative Bloch vector on the dynamics is naturally constrained by 
\eq 
| N_\phi-N_\psi  | \le   |\vec v_{\text{tot}}|  \le N_\phi+N_\psi = N , \label{constraintsvec}
\en
i.e.~in general, the greater the population difference between the two species, the smaller is the influence of the deviation from iso-specificity.

On the other hand, although the amplitude of the perturbation induced by the relative vector in Eq.~(\ref{totalvectordeo}) is constrained by $|(r-1)U N/4|$, a system prepared on the separatrix, i.e.~at the fragile borderline between self-trapping and oscillating regime (see Fig.~\ref{phasespace}), can still be affected qualitatively by very small deviations around $r=1$, which is apparent for $N U/\Omega=2.5$ in Fig.~\ref{MainOverview}(e2): While the singular non-oscillating case is attained for $r=1$, the system is self-trapped for $r=0.95$ and oscillates for $r=1.1$. On the other hand, when the trajectory is stable, and well away from the separatrix, i.e.~deeply in the oscillating or in the self-trapping regime, the change in the dynamics due to the an-isospecificity is very small, as can be observed in Fig.~\ref{MainOverview}(c) and (d) and in the panels (e1,e3,e4), for $U=4, -2,-4$.

Slightly differing inter- and intra-species interaction strengths can lead to chaos at the borderline between self-trapping and oscillation regime \cite{Xu2008}. The effect of exchange symmetry breaking, however, dominates in the regular regime and outweighs the influence of weak an-isospecificity.

\subsection{Experimental implementation}
An experiment that implements a controllable bi-species system can be realized with current technology. A binary system that naturally offers the parameters that we require is given by the $\ket{\psi} :=\ket{F=1,m_F=1}$ and $\ket{\phi} :=\ket{F=1,m_F=-1}$ components of $^{87}$Rb. Since spin-changing collisions are weak \cite{VanKempen2002}, we can assume that the population of either species is conserved and that they can be simultaneously trapped in the same potential \cite{Mele-Messeguer2011}. Moreover, the intra-species interactions are identical, and the ratio of inter- to intra-species interaction fulfills $U_{\psi,\phi}/U_{\psi,\psi} \approx 1.0093 $  \cite{Mele-Messeguer2011,VanKempen2002}.  Depending on the potential difference between the wells at the moment of the cooling, any desired spatial population imbalance can be achieved \cite{Albiez2005}.  An imbalance of the population of the two species can then be implemented selectively on one well by using the techniques utillized in \cite{Weitenberg2011}. Here, a strongly focussed laser beam shifts the hyperfine transition in one well compared to the other. The site-selective transfer is then realized using a microwave pulse at the shifted frequency. 

Using well-established experimental techniques, the total population imbalance can be read off in a destructive way \cite{Albiez2005}; and when combined with a spin-dependent push-out \cite{Weitenberg2011}, the dynamics of the individual species can be observed.

In conclusion, the population of a second species changes the balance between phase-coherence and interaction energy. The new initial state of the system may lie in a possibly different phase-space. For very small ($N U \ll \Omega$) and very large ($N U \gg \Omega$) interactions, the qualitative behavior is  not changed dramatically, and exchange effects are small:  For very weak interactions (in the Rabi regime), each particle tunnels individually, and no collective effects emerge, whereas for very strong interactions, the inter-particle interaction outweights the exchange effects. In typical experimental regimes that lie between these extremes \cite{Albiez2005}, however, the breaking of the exchange symmetry can have a strong qualitative impact on the system dynamics.

\FloatBarrier

\section{Many-body treatment: Spin-coherent states and beyond} \label{BHaappp}
The difference between single- and bi-species condensates can be understood by the competition of phase-coherence and interaction energy. To assess whether the physical arguments can be extended  beyond situations that are well described by mean-field theory, we proceed to an analysis using second quantization. 
\subsection{Two-well Bose-Hubbard model}
The Gross-Pitaevskii equation Eq.~(\ref{eom2}) is the classical limit of the Heisenberg equations of motion retrieved from the Bose-Hubbard Hamiltonian, 
\eq 
H_{\text{BH}}&=&  \frac U 2 \sum_{j=1}^2 ( \hat n_{j,\psi}+\hat n_{j,\phi})( \hat n_{j,\psi}+\hat n_{j,\phi}-1)\nonumber \\
 && - \frac \Omega 2 \left(\hat a^\dagger_{1,\psi} \hat a^{ }_{2,\psi} +\hat a^\dagger_{1,\phi} \hat a^{ }_{2,\phi} + \mbox{H.c.} \right) , \label{BHHamiln} 
\en
where $\hat n_{k,\alpha}$ denotes the number operator, and $\hat a^{\dagger}_{j,\alpha}$ ($\hat a^{\phantom \dagger}_{j,\alpha}$) creates (annihilates) a particle of species $\alpha(=\psi ,\phi)$ in site $j$. Further below, we shall present the results of a numerical simulation of the dynamics induced by this two-species Hamiltonian. In order to investigate the reduction to single-species dynamics, it will be useful to introduce an angular momentum representation, which can be done elegantly for each species using the Schwinger-boson model:  
\eqs
\label{schwinger}
\eq 
\hat J_{z,\alpha}&=&\frac{\hat n_{1,\alpha}-\hat  n_{2,\alpha}} 2 \label{schwinger1} \\ 
\hat J_{x,\alpha}&=&\frac{\hat a^\dagger_{1,\alpha}\hat a^{ }_{2,\alpha} + \hat a^\dagger_{2,\alpha}\hat a^{ }_{1,\alpha} }{2}, \\
 \hat J_{y,\alpha}&=&\frac{i \left( \hat a^\dagger_{2,\alpha}\hat a^{ }_{1,\alpha} - \hat a^\dagger_{1,\alpha}\hat a^{ }_{2,\alpha} \right)}{2} ,\label{schwinger2}
\en
\ens
where $\alpha= \psi ,\phi$. The operators $\hat J_{j,\alpha}$ fulfill the usual SU(2) commutation relations, and the Hamiltonian Eq.~(\ref{BHHamiln}) can be re-written as
\eq 
H_{\text{BH}}&=&   U  \left( \hat J_{z,\psi}+\hat J_{z,\phi} \right)^2 -  \Omega  \left(\hat J_{x,\psi}+\hat J_{x,\phi} \right) , \label{H2spangm}
\en
where a constant summand has been omitted here with respect to Eq.~(\ref{BHHamiln}). From  Eq.~(\ref{H2spangm}), it is immediate that a coupled basis, i.e.~the eigenstates of $\hat J_k= \hat J_{k,\psi}+\hat J_{k,\phi}$  
allows us to re-write the Hamiltonian in the Lipkin-Meshkov-Glick form \cite{LMG1,LMG2,LMG3,Ribeiro2004,Ribeiro2007} 
\eq
H_{\text{BH}} &=&  U  \hat J_{z}^2 - \Omega \hat J_{x}  
  \label{BHHam2} ,
\en
equivalent to the one for a \emph{single} species in the Bose-Hubbard dimer, in direct analogy to Eq.~(\ref{twocomphamil}) -- note that the $z$-projection $m$ is related to $z_{\text{tot}}$ via $2m=z_{\text{tot}}$, which leads to different pre-factors in the consistent Eqs.~(\ref{BHHam2}) and (\ref{twocomphamil}). At this stage, the problem seems to reduce to finding the appropriate single-species configuration, just like in the mean-field representation.  In the discrete case, however, the quality of the single-species reduction depends heavily on the initially chosen configuration, as we shall see. 

\subsection{Bi-species spin-coherent states}
In the mean-field approximation, binomial counting statistics of the many-particle state is assumed \cite{Chuchem2010}, such that the treatment of the last Section is expected to be a good approximation to the spatial population imbalance exhibited by initial states that are spin-coherent, i.e.~of the form 
\begin{widetext}
\eq 
\ket{\Psi_{\text{coh}}}=\frac 1 {\sqrt{N!}} \left[ \sum_{\alpha=\phi,\psi}  \left(\sqrt{ \frac{N_\alpha+z_\alpha}{2 N}}  \hat a_{1,\alpha}^\dagger + e^{i \theta_\alpha} \sqrt{ \frac{N_\alpha-z_\alpha}{2 N}}  \hat a_{2,\alpha}^\dagger   \right) 
   \right]^{N} \ket{\text{vac}} , \label{spincoherentstate}
\en
\end{widetext}
where the number of particles $N$ is large and $\ket{\text{vac}}$ denotes the zero-atom state. The expectation values of this state match precisely the population imbalances $z_\psi$ and $z_\phi$ and the relative phases $\theta_\psi$ and $\theta_\phi$. Note that the relative phase between the species is unobservable in our setting. 

\begin{figure}
\center
\includegraphics[width=.8\linewidth,angle=0]{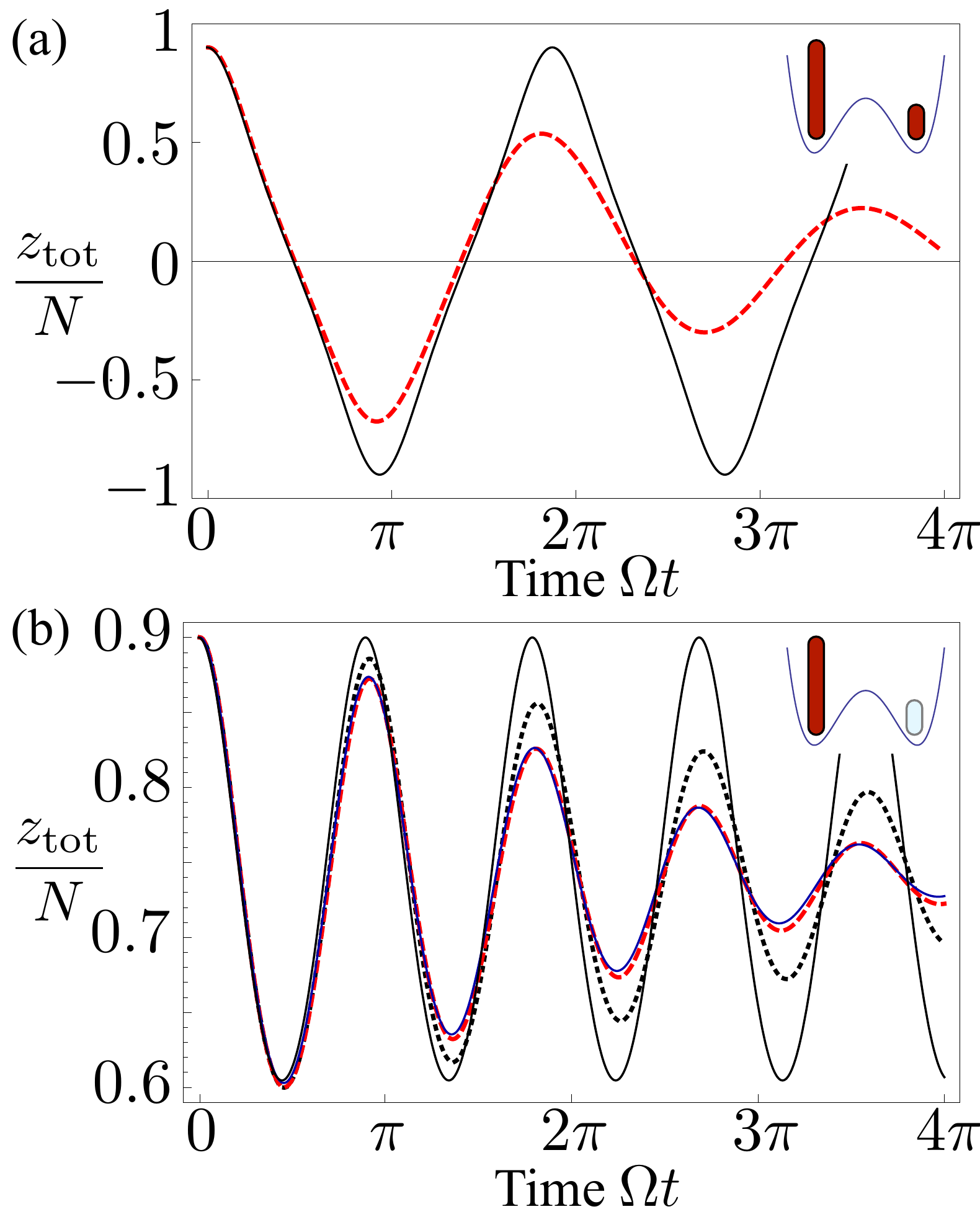} 
\caption{(color online) { Population imbalance: Mean-field treatment and many-body dynamics.} The initial population imbalance is $z_{\text{tot}}(0)=0.9 N$,  $U / \Omega=1/60 $, $N=180$. We compare a spin-coherent many-particle state of the form Eq.~(\ref{spincoherentstate}), for (a) $N_\phi=0$, $N_\psi=N$ (all particles are of the same species) and (b) $N_\phi/N=0.05$, $N_\psi/N=0.95$ (particles in the different wells are of different species) to the classical calculation based on Eq.~(\ref{blocheom2a}). Black solid lines denote the classical calculation, red dashed lines the exact quantum bi-species calculation. The single-species description based on the effective tunneling constant $\Omega$ is shown in blue solid, the one based on the effective total particle number $N_{\text{eff}}=162$ is shown as black dotted line.}
 \label{CountingStatCOH}
\end{figure}

We compare the time-evolution of the population imbalance of a state of the above form Eq.~(\ref{spincoherentstate}) with the classical description based on Eq.~(\ref{blocheom2a}), in Fig.~\ref{CountingStatCOH}. The Hamiltonian Eq.~(\ref{BHHam2}) is integrated in a numerically exact way, without performing any further approximation. The initial spatial imbalance is $z_{\text{tot}}(0)/N=0.9$, the number of particles is $N=180$, and the interaction parameter fulfills $U /\Omega ={1/60}$. We compare (a) the single-species situation to (b) the bi-species scenario. For the latter, we assume that a Rabi $\pi$-pulse on the right well leaves all particles in the second internal state $\phi$, such that $N_\psi/N=0.95$ and $N_\phi/N=0.05$ (see inserted sketches). The classical calculation (black solid line) exhibits oscillations in (a) and self-trapping in (b) (note the changed scale for the ordinate). The quantum many-body calculation features the same qualitative behavior, but the oscillations decay due to quantum fluctuations \cite{Salgueiro2007}. 

The single-species descriptions based on the effective coupling $\bar \Omega$ given in Eq.~(\ref{rescaledOme}) (blue solid line), or employing the effective particle number $N_\text{eff}$ given in Eq.~(\ref{classicaleffpn}) (black dotted line), reproduce the dynamics well, and also feature the decay of  oscillations. In particular, switching the population of the right well to the second species by a $\pi$ Rabi pulse clearly switches the behavior of the system from Josephson-oscillations to self-trapping - just like in Fig.~\ref{MainOverview}. The decay of the oscillations for the description based on the effective particle number $N_\text{eff}=162$ (black dotted) is slower than for the two-species scenario (red dashed) and for the description via the effective tunneling coupling $\bar \Omega$ (blue solid), since all $N_{\text{eff}}$ particles start in the left well, which reduces quantum fluctuations.

\subsection{Bi-species Fock-states}
Since relative phase and particle number difference are conjugated variables, the phase relation between two wells populated by the same species can also be broken by preparing a Fock-state with a well-specified number of particles in each well.  In the semiclassical picture, a Fock-state corresponds to a distribution on the Bloch-sphere with well-specified $z$-component, but uncertain $x$- and $y$-component, i.e.~to a ring on the Bloch-sphere, or a vertical line in Fig.~\ref{phasespace}. The resulting distributions for single-species and bi-species Fock-states differ, and so does the dynamics, as we will see below. 

\subsubsection{Single-species description}
We now derive an effective single-species description of a bi-species Fock-state, analogous to Section \ref{amplitudescaling}, and discuss the requirements for its application. 
\begin{figure*}
\center
\includegraphics[width=\linewidth,angle=0]{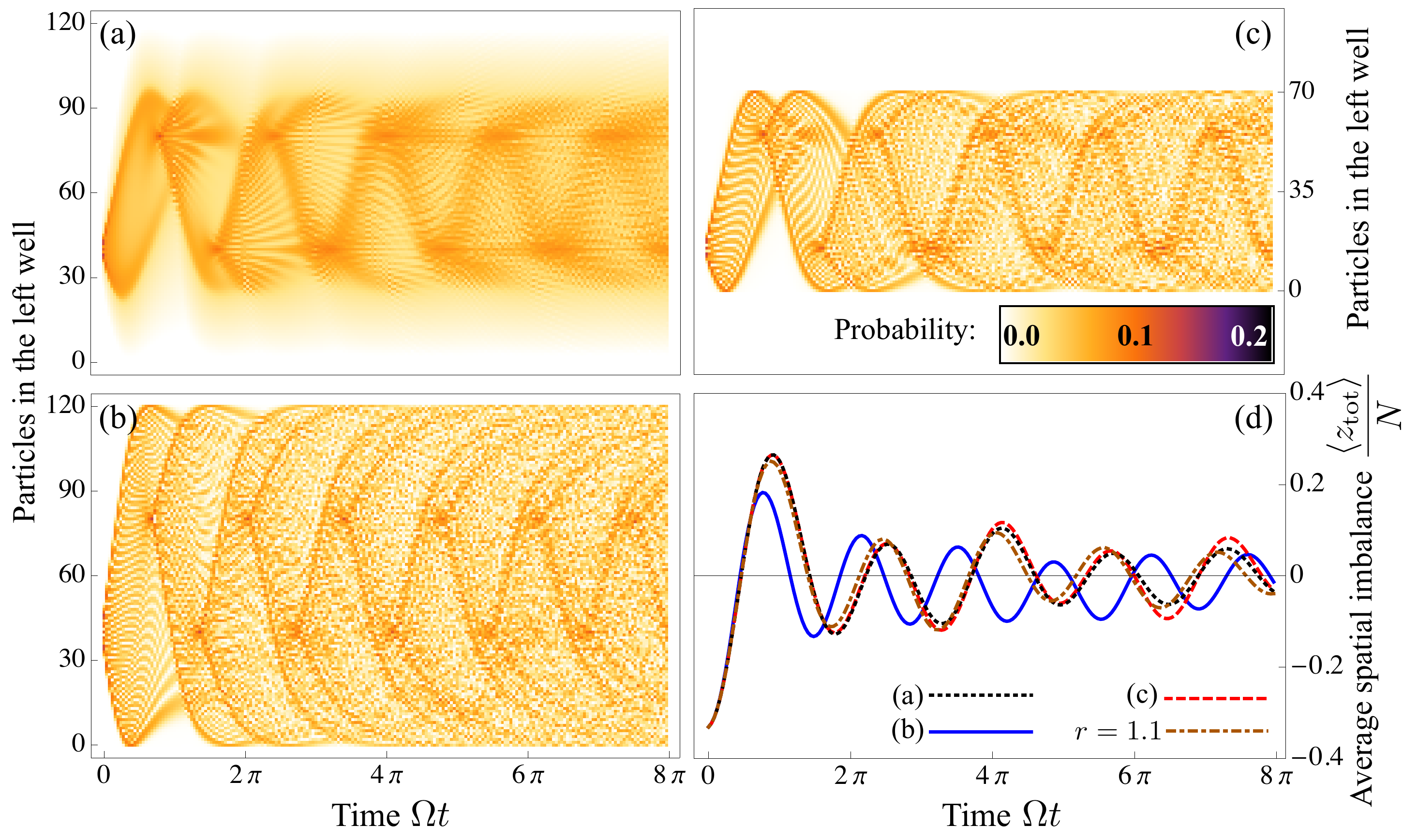} 
\caption{(color online) { Time-evolution of Fock-states for $N=120$, $U/\Omega=0.021$.} (a,b) Counting statistics for single- and bi-species Fock-state with initially 40 (80) particles in the left (right) well. Horizontal axis: Time $t$. Vertical axis: Number of particles in the left well. The color-code indicates the probability. (a) Bi-species Fock-state Eq.~(\ref{bisp1}) with $N_\psi \equiv 2j_\psi=60, N_\phi \equiv 2 j_\phi=60, m_\psi=10, m_\phi=-j_\phi=-30$, $m_{\text{tot}}=-20$. (b) Single-species state Eq.~(\ref{bisp2})  with $N_\psi=120, m_\psi=-20$. (c) Single-species approximation  Eq.~(\ref{bisp3})  for the bi-species state (a): $N_{\text{eff}}=70, m=-20$. (d) Average population imbalance, for the bi-species Fock-state (black dotted), single-species Fock-state (blue solid), the single-species approximation (red dashed) and the weakly an-isospecific case $r=1.1$ (brown dot-dashed).} 
 \label{CountingStat}
\end{figure*}
The Hamiltonian Eq.~(\ref{BHHam2}) suggests to work in the basis of \emph{coupled} angular momenta, i.e.~in eigenstates of $\hat J^2, \hat J_\psi^2, \hat J_\phi^2, \hat J_z$, 
\eq 
\ket{j, j_\psi, j_\phi, m} , \label{totaljeigenst}
\en
where $j_\psi=N_\psi/2$ and $j_\phi=N_\phi/2$ are related to the number of particles of either species, and $m$ is the total spatial particle number imbalance. The quantum number $j$ as well as the total particle number of either species, $N_\psi$ and $N_\phi$,  are constants of motion -- these conserved quantities are the analogous quantities to $|\vec v_{\text{tot}}|, |\vec  v_\psi|$, and $|\vec  v_\phi|$ in the classical description.

An initial Fock-state $\ket{\Psi_{\text{F}}}$ with a well-defined number of particles of either species in each well is, in general, not an Eigenstate of $\hat J^2$ of the form Eq.~(\ref{totaljeigenst}), but an eigenstate of the uncoupled angular momentum operators $\hat J^2_\psi, \hat J_{z,\psi},\hat J^2_\phi, \hat J_{z,\phi}$. It can, however, be written as a superposition of total angular momentum eigenstates given in Eq.~(\ref{totaljeigenst}), 
\eq
\ket{\Psi_{\text{F}}}=\ket{j_\psi, m_\psi, j_\phi, m_\phi }=\sum_{j=|j_\psi-j_\phi|}^{j_\psi+j_\phi} c_{j}  \ket{j, j_\psi, j_\phi, m} , \label{ClebschGordan}
\en
where $m=m_\psi+m_\phi$ and $c_j$ is the Clebsch-Gordan coefficient \eq c_j= \braket{j, j_\psi, j_\phi, m}{j_\psi, m_\psi, j_\phi, m_\phi }. \en

The single-species description with a fixed effective total particle number corresponds to the approximation 
\eq 
\ket{\Psi_{\text{F}}}=\ket{j_\psi, m_\psi, j_\phi, m_\phi} \approx \ket{j_{\text{eff}}, j_\psi, j_\phi, m_\phi+m_\psi} , \label{approxsinglespec}
\en
where $j_{\text{eff}}$ is the expectation value of the total angular momentum, \eq j_{\text{eff}} (j_{\text{eff}}+1) &=& \bra{\Psi_{\text{F}}}  \hat J^2 \ket{\Psi_{\text{F}}}   \label{relationj}   \\
 &=& j_\psi(j_\psi+1)+j_\phi(j_\phi+1) + 2 m_\psi m_\phi , \nonumber \en
 which allows an interpretation as effective particle number, $N_{\text{eff}}/2=j_{\text{eff}}$. The classical limit of Eq.~(\ref{relationj}) is indeed Eq.~(\ref{classicaleffpn}), which neglects linear terms with respect to Eq.~(\ref{relationj}). 

The approximation Eq.~(\ref{approxsinglespec}) is reasonable only for states $\ket{\Psi_{\text{F}}}$ that possess a well-defined total angular momentum, i.e.~when 
\eq
\frac{ \Delta (\hat J^2) } { \langle \hat J^2 \rangle } \ll 1 , \label{deltaj2}
\en
such that the distribution of Clebsch-Gordan coefficients $c_j$ that appears in Eq.~(\ref{ClebschGordan}) is narrow and peaked at $j_{\text{eff}}$. The variance of the expectation value of $\vec J^2$ amounts to 
\eq 
\Delta \left(\hat J^2\right) &=& \sqrt{2(j_\psi + j_\psi^2-m_\psi^2)(j_\phi + j_\phi^2-m_\phi^2) -2 m_\phi m_\psi } \nonumber ,
\en
i.e.~the uncertainty in the total angular momentum depends on the projection $m_\phi$, $m_\psi$ of the two angular momenta. Therefore, depending on the physical situation, the relation Eq.~(\ref{approxsinglespec}) is exact, approximate, or unsuitable in the limit of many particles:
\begin{itemize}
\item {\bf Only one well occupied.} We then have $m_\psi=j_\psi, m_\phi=j_\phi$ and find the \emph{equality} 
\eq 
 \ket{j_\psi, m_\psi=j_\psi, j_\phi, m_\phi=j_\phi} = \nonumber \hspace{1cm} \\ \ket{j = j_\psi+ j_\phi , j_\psi,  j_\phi, m=m_\psi + m_\phi} .
\en
Since all particles are prepared in the same well, the above state is  fully exchange-symmetric and behaves like a Fock-state of a single species. 
\item {\bf One species localized in one well.} If all particles of one species are localized in one well while the particles of the other species remain  distributed among the two wells, i.e.~without restrictions of generality $m_\phi=j_\phi$, the relative variance Eq.~(\ref{deltaj2}) becomes small in the limit of large total particle numbers -- we assume that $N$ is increased for a constant ratio of species populations and constant population imbalances (constant $m_\psi/j_\psi$, $m_\phi/j_\phi$,  and $j_\phi/j_\psi$). Then, Eq.~(\ref{approxsinglespec}) is approximately valid. The scaling property Eq.~(\ref{effectivecoupling}), which was shown to hold for spin-coherent states, is here, however, not applicable anymore: In the derivation Eq.~(\ref{rescaledOme}) of $\bar \Omega$, we assumed a fixed value of $\theta_\phi$ and $\theta_\psi$. Since the initial phase relationship is uncertain -- a Fock-state is prepared -- no effective tunneling $\bar \Omega$ emerges. 
\item {\bf General state.} If both wells are occupied by particles of either species, we have $|m_\psi|<j_\psi$ and $|m_\phi|<j_\phi$. The distribution of Clebsch-Gordan coefficients in Eq.~(\ref{ClebschGordan}) remains broad for increased total particle number, and, in particular, the distribution of the $c_j$ is not necessarily peaked at $j_{\text{eff}}$. The relative variance Eq.~(\ref{deltaj2}) does not vanish for large particle numbers, and the approximation Eq.~(\ref{approxsinglespec}) is unsuitable. 
\end{itemize}

We exemplify the validity of the approximation Eq.~(\ref{approxsinglespec}) for a two-species Fock-state in Fig.~\ref{CountingStat}, for $U/\Omega =0.021$, $N=120$. Panels (a-c) show the probability for a certain number of particles in the left well, and panel (d) compares the average population imbalance obtained by the single- and two-species numerical solution and by the approximation. In (a), the initial state is 
\eq \ket{\Psi_{\text{F,2}}}=\ket{j_{\psi}=30, m_\psi=10, j_\phi=30, m_\phi=-30} ,  \label{bisp1} \en
which leads to a very distinct counting statistics when compared to a single-species state with the same initial spatial population imbalance in (b),
\eq 
\ket{\Psi_{\text{F,1}}}=\ket{j_{\psi}=60, m_\psi=-20} .  \label{bisp2}
\en
In particular, the probability to find almost all or almost no particle in one mode is clearly enhanced for $\ket{\Psi_{\text{F,1}}}$ with respect to $\ket{\Psi_{\text{F,2}}}$, well in accordance with our intuition for bosons. The typical features of the counting statistics of $\ket{\Psi_{\text{F,2}}}$ are well reproduced by the single-species approximation based on Eq.~(\ref{approxsinglespec}), 
\eq 
\ket{\Psi_{\text{F,1,appr.}}}=\ket{j_{\text{eff}}=35, m_{\text{}}=-20} ,  \label{bisp3}
\en
where the vertical axis of panel (c) is displaced for direct comparison to (a) and (b). Not only is the suppression of events with many particles in the left or right well and the overall structures of the counting statistics reproduced, also the characteristic oscillation of the center-of-mass agrees very well, as can be observed in (d). The more fine-grained interference pattern in (c) with respect to (a) can be understood from the unique contribution of one term in the sum Eq.~(\ref{ClebschGordan}), i.e.~for $\ket{\Psi_{\text{F,2}}}$, many-particle interference is averaged out.

\begin{figure}
\center
\includegraphics[width=.85\linewidth,angle=0]{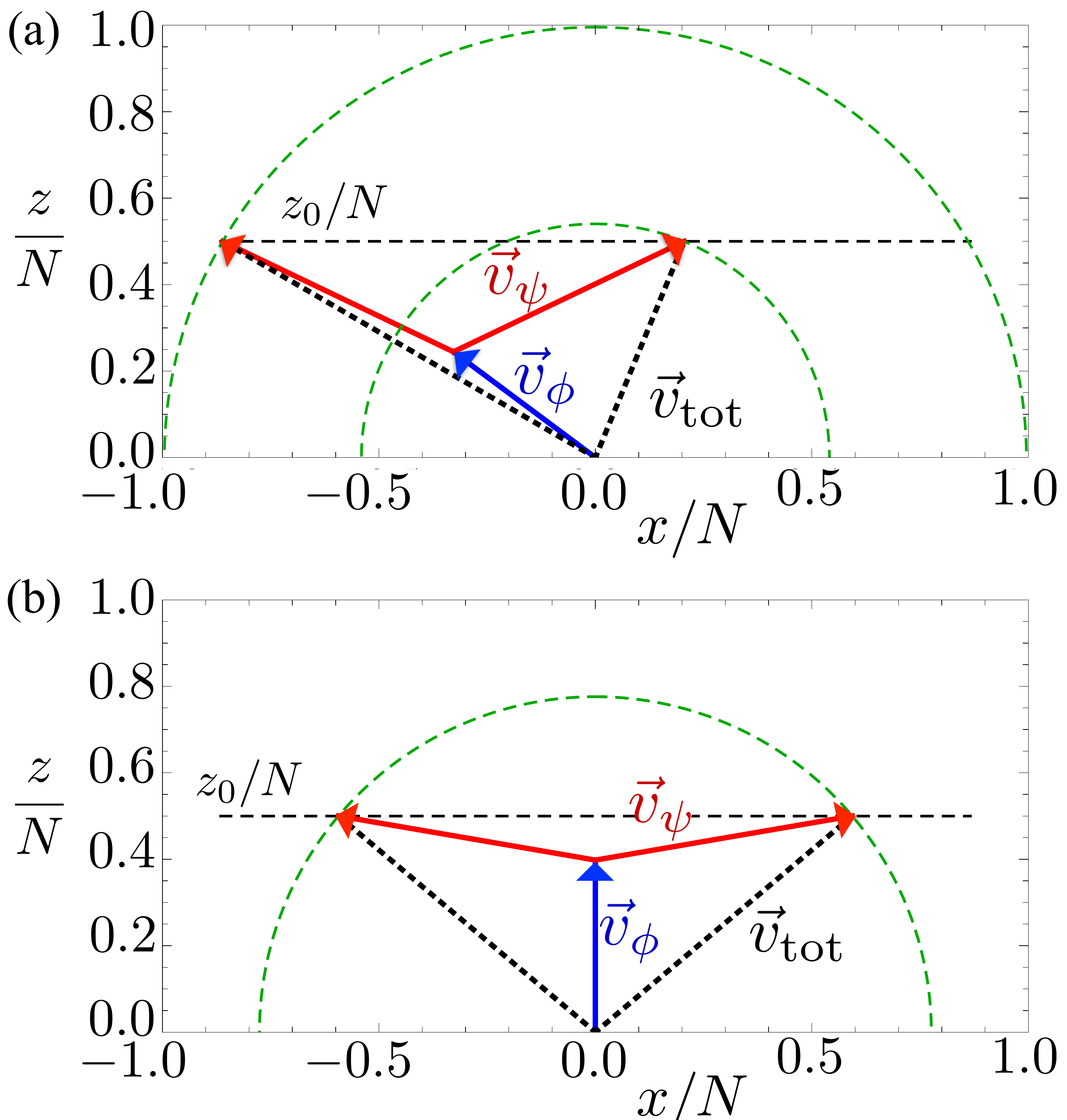} 
\caption{(color online) { Coupling of Bloch-vectors with specified $z$-component, but random phase.} We show a cut through the Bloch-sphere, for $y=0$. Since the relative phases between the wells are unknown for either species, the length of the emerging vector $\vec v_{\text{tot}}$ is also uncertain. In  (a), $\vec v_{\psi}$ and $\vec v_\phi$ couple to a total vector $\vec v_{\text{tot}}$ with a length ranging between the small and the large dashed-green Bloch-sphere, since only the $z$-component, but not the $x-$ and $y-$ component of the two species are fixed. The vector $\vec v_\phi$ is fixed in the representation for convenience. The effective total particle number is not well defined, but follows a distribution that corresponds to the classical limit of the Clebsch-Gordan coefficients. In (b), the $x$- and $y$- components of the $\phi$-species vanishes, which also fixes the length of $\vec v_{\text{tot}}$. In the corresponding many-body situation, the approximation Eq.~(\ref{approxsinglespec}) is valid for large particle numbers.}  
 \label{blochvectorcouplingfigsemiclas}
\end{figure}
\subsubsection{Semiclassical description}
An intuitive argument for the validity of the single-species approximation Eq.~(\ref{approxsinglespec}) for  $m_\phi=\pm j_\phi$ can be obtained in a semi-classical picture. A Fock-state corresponds to a Bloch-vector with fixed $z$-component, but uncertain $x$- and $y$-component -- the phase between the wells is maximally uncertain, since the particle number in each well is fixed \cite{Chuchem2010,Smith-Mannschott2009,Mullin2006,Mullin2010}. A single-species Fock-state is thus described by a vector of constant length that lies on a cone around the z-axis. A bi-species Fock-state corresponds then to two coupled vectors, $\vec v_\psi$ and $\vec v_\phi$, of fixed length and fixed $z$-component, but uncertain $x$- and $y$-components. The $z$-component of the coupled, total vector is well-specified, $z_{\text{tot}}=z_{\phi}+z_{\psi}$. In general, however, the \emph{length} of the total Bloch-vector is not fixed, since the relative orientation of $\vec v_\psi$ and $\vec v_\phi$ is uncertain. The bi-species system behaves like a single-species system with a total particle number that is not fixed, but which follows a certain probability distribution, $P(|\vec v_{\text{tot}} |) $, which is precisely the classical analogous \cite{Coefficients} to the Clebsch-Gordan expansion Eq.~(\ref{ClebschGordan}). This situation is also depicted in Fig.~\ref{blochvectorcouplingfigsemiclas} (a).

When one of the vectors, say $\vec v_\phi$, points to the $z$-direction, it then exhibits vanishing uncertainty in the $x$- and $y$-components -- the cone on which it rotates shrinks to a point. Consequently, also $\vec v_{\text{tot}}$ moves on a cone and has a constant, well-defined length, as can be observed in Fig.~\ref{blochvectorcouplingfigsemiclas}(b).

\subsubsection{Robustness against an-isospecificity}
Again, the question arises how robust the dynamics are with respect to weakly an-isospecific interactions, i.e.~when $r\neq 1$ and thus $U_{\psi,\phi} \neq U_{\psi,\psi}$ (see Eq.~(\ref{isospecificityU})). Due to the Heisenberg uncertainty relation, and due to $z$ and $\theta$ being conjugate variables, no quantum state can be prepared precisely on the separatrix between self-trapping and oscillation regime. Thus, the pathological case shown in Fig.~\ref{MainOverview}(e2) has no analogy in the quantum treatment. 

  Since Fock-states can be described semiclassically as a distribution of classical solutions with fixed population imbalance $z$ and  unknown phase $\theta$, the overall behavior of the system is seldom clearly in the self-trapping or in the oscillation regime (consider a distribution shaped as a vertical line in Fig.~\ref{phasespace}), and thus widely unaffected by small changes in $r$. We show the average spatial imbalance for $r=1.1$ in Fig.~\ref{CountingStat} (d), it remains close to the behavior of the system with $r=1$ (a). 

Summarizing, just like for the mean-field treatment, the impact of the exchange symmetry breaking can change the system behavior qualitatively, while inter- to intra-species interaction ratios that deviate little from unity do not jeopardize the general system behavior.

\FloatBarrier

\section{Effective single-species description of general multi-species Bose-Hubbard models} \label{generalized}
In the two previous sections, we have shown that the total particle density in a mixture can be be expressed as the density of an effective single species, either exactly -- in the classical treatment -- or approximately -- in the quantum many-body treatment. On the one hand, this allows us to understand the effect of a broken exchange symmetry, on the other hand, especially in the many-body treatment, the relationship allows to save significant computational resources \cite{Venzl2009}. It is therefore appealing to expand the discussion to more general scenarios. 

Whereas the non-linear scaling of the initial population imbalance, Eq.~(\ref{scalingprop}), is a peculiarity of the two-mode model, we can generalize the argument of the effective particle number to an arbitrary number of species and wells. 

\subsection{Many-site Bose-Hubbard model}
For this purpose, we consider a general $L$-site Bose-Hubbard Hamiltonian that describes $s$ distinct species, 
\eq 
\hat H_{\text{BH,gen}} &=& - \frac 1 2 \sum_{j>k}^L \Omega_{j,k} ~\left( \sum_{\alpha=1}^s \hat a_{j,\alpha}^\dagger \hat a^{\phantom \dagger }_{k,\alpha}+\hat a_{j,\alpha}^{\phantom \dagger} \hat a^{\dagger }_{k,\alpha} \right) \nonumber \\ && 
+  \sum_{l=1}^L \epsilon_l \left( \sum_{\alpha=1}^s\hat n_{l,\alpha}  \right) \nonumber \\ 
&& + \frac U 2 \sum_{j=1}^L \left( \sum_{\alpha=1}^s \hat n_{j,\alpha} \right)^2  , \label{hamilgen}
\en
where $\epsilon_j$ is the on-site energy for the $j$-th site, $\Omega_{j,k}$ is the tunneling coupling between site $j$ and $k$, $U$ is the interaction strength. The creation operatore $\hat a_{j,\alpha}^\dagger$ creates a particle of species $\alpha$ on site $j$. The Hamiltonian Eq.~(\ref{hamilgen}) is isospecific, just like Eq.~(\ref{BHHamiln}). 

In analogy to the Schwinger-Boson algebra, which allows to rewrite the two-site Hamiltonian in terms of the three generators of SU(2) (see Eqs.~(\ref{schwinger}), (\ref{H2spangm})), we can re-write Eq.~(\ref{hamilgen}) in terms of the $L^2-1$  generators of SU($L$), for each species $\alpha$. This approach has been applied to the triple-well case \cite{Franzosi2001,Viscondi2010,Viscondi2011} and can be readily generalized to the present $L$-site system with $s$ species. 

The $L-1$ diagonal generators of SU($L$) can be chosen as
\eq
\hat Z^{(k)}_\alpha=\frac{\sum_{p=1}^k \hat n_{p,\alpha} - k \cdot \hat n_{k+1,\alpha} }{k+1},
\en
where $k \in \{ 1, \dots, L-1\}$, while the $L(L-1)$ hopping operators read
\eq
\hat X^{(l,m)}_\alpha &=& \frac{\hat a_{l,\alpha}^\dagger  \hat a_{m,\alpha}^{\phantom \dagger} + \hat a_{l,\alpha}^{\phantom \dagger}  \hat a_{m,\alpha}^{\dagger} }{2}, \\ 
\hat Y^{(l,m)}_\alpha &=& \frac{ i \left( \hat a_{l,\alpha}^\dagger  \hat a_{m,\alpha}^{\phantom \dagger} -  \hat a_{l,\alpha}^{\phantom \dagger}  \hat a_{m,\alpha}^{\dagger}\right)}{2}, 
\en
where $l > m$. For $L=2$, we simply retrieve Eq.~(\ref{schwinger}). 

For a given species $\alpha$, the $L$ operators $\hat n_{j,\alpha}$ can thus be expressed by the $L-1$ operators $Z^{(k)}_\alpha$ and the total particle number operator $\hat N_\alpha=\sum_{l=1}^L \hat n_{l,\alpha}$:
\eq 
\hat n_{j,\alpha} =\frac{\hat N}{L} - \hat Z^{(j-1)}_\alpha  + \sum_{k=j}^{L-1}  \frac{\hat Z^{(k)}_\alpha}{k} ,
\en
where we set $\hat Z^{(0)}_\alpha=0$ for convenience. 

The SU($L$) commutation relations of the $\hat Z^{(k)}_\alpha$, $\hat X^{(l,m)}_\alpha$ and $\hat Y^{(l,m)}_\alpha$ are immediately inherited by the coupled operators
\eq 
\hat Z^{(k)} &=& \sum_{\alpha=1}^s \hat Z^{(k)}_\alpha ,   \\  
 \hat X^{(l,m)}&=& \sum_{\alpha=1}^s \hat X^{(l,m)}_\alpha , \ \ \ \ 
  \hat Y^{(l,m)}= \sum_{\alpha=1}^s \hat Y^{(l,m)}_\alpha ,\nonumber
\en
since -- just like in the two-site case -- operators related to different species always commute. 

The Hamiltonian Eq.~(\ref{hamilgen}) can then be expressed with these coupled operators, 
\eq 
\hat H_{\text{BH,gen}} &=&  - \sum_{j>k}^L  \Omega_{j,k} \hat X^{(j,k)} \label{SULHamiltonian} \\ 
&&+\sum_{k=1}^{L-1} \hat Z^{(k)}  \left( \sum_{p=1}^k \frac{\epsilon_p}{k} - \epsilon_{k+1} \right) + \hat N \bar \epsilon \nonumber \\
 &&+ \frac U 2 \left( \frac{\hat N^2}{L}+ \sum_{k=1}^{L-1} \frac {(k+1) } {k}  \left( \hat Z^{(k)} \right)^2 \right) \nonumber  ,
\en
where $\hat N$ is the total particle number, and $\bar \epsilon = (\sum_{k=1}^L \epsilon)/L$ the average onsite-energy. The constant terms proportional to $\hat N$ and $\hat N^2$ can be neglected for the dynamics. Just like for $L=2$, the many-species Hamiltonian assumes the same form as for a single species. Again, any multi-species state can be expanded in single-species states: A many-particle state of $N_\alpha$ particles is defined  by $L$ quantum numbers, i.e.~a general Fock-state with a well-defined number of particles in each mode is described by
\eq 
\ket{\Psi_{\text{F,gen}}}=\otimes_{\alpha=1}^s \ket{N_\alpha, m_{\alpha, 1}, m_{\alpha, 2}, \dots , m_{\alpha, {L-1}} } ,
\en
where $m_{\alpha,k}$ is the quantum number  corresponding to the $\hat Z^{(k)}_\alpha$ operator and $N_\alpha$ is the total particle number of the species $\alpha$. In analogy to Eq.~(\ref{ClebschGordan}), $\ket{\Psi_{\text{F,gen}}}$ can be expanded in a sum of total particle number states, 
\eq 
\ket{\Psi_{\text{F,gen}}}= \hspace{6cm} \\ \nonumber  \sum_{N_{\text{tot}} } c_{N_{\text{tot}}} \ket{N_{\text{tot}}, N_1, \dots , N_s, \sum_{\alpha=1}^s m_{\alpha,1}, \dots,  \sum_{\alpha=1}^s m_{\alpha,L-1} } , 
\en
where the $c_{N_{\text{tot}}}$ are the generalized Clebsch-Gordan coefficients for SU($L$) \cite{Alex2011}.

\subsection{Classical limit}
The Hamiltonian Eq.~(\ref{SULHamiltonian}) allows us to retrieve the Heisenberg equations of motion for the $s(L^2-1)$ time-dependent operators in the Heisenberg picture, exploiting the commutation relations of the generators of SU($L$). In the mean-field limit, these equations become discrete Gross-Pitaevskii equations for vectors $\vec v_\alpha$ on $s$ $(L^2-1)$-dimensional Bloch-spheres, analogous to Eq.~(\ref{blocheom2a}):
 \eq
 \frac{\text{d}}{\text{d}t} \vec v_\alpha = M\left( \sum_{\beta=1}^s \vec v_\beta \right) \vec v_{\alpha} ,
 \en 
 where $M$ is a matrix of dimension $(L^2-1)\times (L^2-1)$. Since $M$ depends only on the total Bloch-vector
 \eq 
 \vec v_{\text{tot}} = \sum_{\alpha=1}^s \vec v_\alpha,
 \en
the latter fulfills
\eq
 \frac{\text{d}}{\text{d}t} \vec v_{\text{tot}} = M\left(  \vec v_{\text{tot}} \right)  \vec v_{\text{tot}} .
 \en
That is to say, the inclusion of a new species with the same physical properties as the existing one leads, again, to a total Bloch-vector $\vec v_{\text{tot}}$, whose length can be interpreted as an effective particle number. 
\FloatBarrier

\section{Conclusions} \label{Conclusions}
Ultracold atoms in optical lattices do not only constitute ideal model systems to study the effects of tunable  inter-particle interaction, e.g.~via Feshbach resonances \cite{Bloch2008}, but also to investigate the consequences of a broken exchange symmetry or \emph{distinguishability} of particles via the population of different hyperfine states. Thereby, the interplay of interaction and exchange symmetry -- the two pertinent ingredients of quantum many-body dynamics -- may be explored. We investigated a two-well system, in which the population of a second internal state effectively breaks the phase relationship between the macroscopic wavefunction in the two wells, and thus influences the dynamics of the system.  The bosonic Josephson junction exhibits symmetries that allow us to understand the dynamics either by an effective tunneling coupling, or by an effective total particle number. 

For double-Fock-states, for which no phase relation between the wells exists, exchanging particles of one species by another breaks the exchange symmetry of the total many-particle state. The effective particle number is then, in general, not constant, but follows a probability distribution, which affects the many-particle interference-capability of the system, and which can be observed  in the emerging counting statistics. 

We showed that controlling a degree of freedom that is invisible for the Hamiltonian of a system can effectively steer and influence the dynamics considerably by breaking or restoring phase coherence. 
Here, the dynamics in the double-well can be switched between the oscillatory and the self-trapping regime, which may be used as an effective interaction-controlling parameter, e.g.~when Feshbach resonances are unaccessible. While we have confined ourselves to the comparison of situations with constant species population, \emph{driving} a system via breaking and restoring its  phase coherence may allow for further control \cite{Zhang2012}.

By relaxing the assumption of perfect isospecificity, we also showed that the breaking of the exchange symmetry has a much greater impact on the system dynamics than weakly differing inter- and intra-species interactions. Only in the fragile case of orbits near the transition between dynamical regimes does a small change in inter- to intra-particle interaction strength jeopardize the single-species model.  With the inclusion of more species simulations become more and more demanding, and a single-species approximation can potentially offer a significant computational advantage. It remains to be studied to which extent quantum phases of multicomponent bose gases can be described by methods analogous to those exposed here.

\subsubsection*{Acknowledgements}  M.C.T. would like to thank Marina Mel\'e-Messeguer for very helpful discussions, and gratefully acknowledges support by the Alexander von Humboldt-Foundation through a Feodor Lynen Fellowship.

\end{document}